\documentclass[aps,pra,10pt,twocolumn,floatfix,nofootinbib,superscriptaddress,
longbibliography,nobibnotes]{revtex4-2}

\usepackage{amsmath}
\usepackage{mathtools}
\usepackage{wasysym}
\usepackage{newtxmath}
\usepackage{mathrsfs}

\usepackage{array} 
\usepackage[T1]{fontenc}
\usepackage{newtxtext}
\usepackage{dsfont}                 %
\usepackage{bbold}                  %
\usepackage[normalem]{ulem}         %

\usepackage{enumerate}
\usepackage[shortlabels]{enumitem}

\usepackage[dvipsnames,table]{xcolor}
\usepackage{graphicx}
\graphicspath{{figs/}}              %

\usepackage[flushleft]{threeparttable}
\usepackage{multirow}

\usepackage{physics}                %
\usepackage{csquotes}               %
\usepackage{comment}

\usepackage[caption=false]{subfig}
\usepackage[colorlinks=true,citecolor=cyan,linkcolor=magenta,filecolor=magenta]{hyperref}
\usepackage[capitalize]{cleveref}

\usepackage{tikz}
\usetikzlibrary{decorations.pathmorphing}
\usetikzlibrary{arrows,decorations.markings,cd}

\usepackage{orcidlink}

\DeclareMathAlphabet{\mathbbold}{U}{bbold}{m}{n}

\newcommand{\be}{\begin{equation}}
\newcommand{\ee}{\end{equation}}

\def\bra#1{\langle{#1}|}				%
	\def\ket#1{|{#1}\rangle}		%

\usepackage{bm}

\definecolor{TB}{rgb}{0.93,0.47,0.2}

\newcommand{\OBCtikz}{%
\begin{tikzpicture}[x=0.75pt,y=0.75pt,yscale=-1,xscale=1,scale=0.6]
				\draw  [color={rgb, 255:red, 0; green, 0; blue, 0 }  ,draw opacity=1 ][fill={rgb, 255:red, 255; green, 255; blue, 255 }  ,fill opacity=1 ]  (211.26,-2.77) -- (385.41,-2.77) -- (385.41,30.46) -- (211.26,30.46) -- cycle  ;
				\draw  [fill={rgb, 255:red, 155; green, 155; blue, 155 }  ,fill opacity=1 ] (219.33,10.83) .. controls (219.33,7.98) and (221.65,5.67) .. (224.5,5.67) .. controls (227.35,5.67) and (229.67,7.98) .. (229.67,10.83) .. controls (229.67,13.69) and (227.35,16) .. (224.5,16) .. controls (221.65,16) and (219.33,13.69) .. (219.33,10.83) -- cycle ;
				\draw  [fill={rgb, 255:red, 155; green, 155; blue, 155 }  ,fill opacity=1 ] (256.33,10.67) .. controls (256.33,7.81) and (258.65,5.5) .. (261.5,5.5) .. controls (264.35,5.5) and (266.67,7.81) .. (266.67,10.67) .. controls (266.67,13.52) and (264.35,15.83) .. (261.5,15.83) .. controls (258.65,15.83) and (256.33,13.52) .. (256.33,10.67) -- cycle ;
				\draw    (229.67,10.83) -- (256.33,10.67) ;
				\draw [shift={(247.8,10.72)}, rotate = 179.64] [color={rgb, 255:red, 0; green, 0; blue, 0 }  ][line width=0.75]    (8.74,-3.92) .. controls (5.56,-1.84) and (2.65,-0.53) .. (0,0) .. controls (2.65,0.53) and (5.56,1.84) .. (8.74,3.92)   ;
				\draw  [fill={rgb, 255:red, 155; green, 155; blue, 155 }  ,fill opacity=1 ] (293,10.67) .. controls (293,7.81) and (295.31,5.5) .. (298.17,5.5) .. controls (301.02,5.5) and (303.33,7.81) .. (303.33,10.67) .. controls (303.33,13.52) and (301.02,15.83) .. (298.17,15.83) .. controls (295.31,15.83) and (293,13.52) .. (293,10.67) -- cycle ;
				\draw    (266.33,10.83) -- (293,10.67) ;
				\draw [shift={(284.47,10.72)}, rotate = 179.64] [color={rgb, 255:red, 0; green, 0; blue, 0 }  ][line width=0.75]    (8.74,-3.92) .. controls (5.56,-1.84) and (2.65,-0.53) .. (0,0) .. controls (2.65,0.53) and (5.56,1.84) .. (8.74,3.92)   ;
				\draw  [fill={rgb, 255:red, 155; green, 155; blue, 155 }  ,fill opacity=1 ] (330,10.33) .. controls (330,7.48) and (332.31,5.17) .. (335.17,5.17) .. controls (338.02,5.17) and (340.33,7.48) .. (340.33,10.33) .. controls (340.33,13.19) and (338.02,15.5) .. (335.17,15.5) .. controls (332.31,15.5) and (330,13.19) .. (330,10.33) -- cycle ;
				\draw    (303.33,10.5) -- (330,10.33) ;
				\draw [shift={(321.47,10.39)}, rotate = 179.64] [color={rgb, 255:red, 0; green, 0; blue, 0 }  ][line width=0.75]    (8.74,-3.92) .. controls (5.56,-1.84) and (2.65,-0.53) .. (0,0) .. controls (2.65,0.53) and (5.56,1.84) .. (8.74,3.92)   ;
				\draw  [fill={rgb, 255:red, 155; green, 155; blue, 155 }  ,fill opacity=1 ] (367.67,10.67) .. controls (367.67,7.81) and (369.98,5.5) .. (372.83,5.5) .. controls (375.69,5.5) and (378,7.81) .. (378,10.67) .. controls (378,13.52) and (375.69,15.83) .. (372.83,15.83) .. controls (369.98,15.83) and (367.67,13.52) .. (367.67,10.67) -- cycle ;
				\draw    (341,10.83) -- (367.67,10.67) ;
				\draw [shift={(359.13,10.72)}, rotate = 179.64] [color={rgb, 255:red, 0; green, 0; blue, 0 }  ][line width=0.75]    (8.74,-3.92) .. controls (5.56,-1.84) and (2.65,-0.53) .. (0,0) .. controls (2.65,0.53) and (5.56,1.84) .. (8.74,3.92)   ;

				\draw  [color={rgb, 255:red, 0; green, 0; blue, 0 }  ,draw opacity=1 ][fill={rgb, 255:red, 255; green, 255; blue, 255 }  ,fill opacity=1 ]  (349.6,21.43) -- (490.74,21.43) -- (490.74,57.83) -- (349.6,57.83) -- cycle  ;
				\draw (356.73,25.24) node [anchor=north west][inner sep=0.75pt]  [font=\small,rotate=-0.25,scale=0.9]  {$G_{\operatorname{OBC}}( z,\overline{z}) =\frac{1}{z}$};
\end{tikzpicture}%
}

\newcommand{\PBCtikz}{%
\begin{tikzpicture}[x=0.75pt,y=0.75pt,yscale=-1,xscale=1,scale=0.6]

                \draw  [color={rgb, 255:red, 0; green, 0; blue, 0 }  ,draw opacity=1 ][fill={rgb, 255:red, 255; green, 255; blue, 255 }  ,fill opacity=1 ]  (191.26,-2.77) -- (405.41,-2.77) -- (405.41,38.46) -- (191.26,38.46) -- cycle  ;

				\draw  [fill={rgb, 255:red, 155; green, 155; blue, 155 }  ,fill opacity=1 ] (219.33,10.83) .. controls (219.33,7.98) and (221.65,5.67) .. (224.5,5.67) .. controls (227.35,5.67) and (229.67,7.98) .. (229.67,10.83) .. controls (229.67,13.69) and (227.35,16) .. (224.5,16) .. controls (221.65,16) and (219.33,13.69) .. (219.33,10.83) -- cycle ;
				\draw  [fill={rgb, 255:red, 155; green, 155; blue, 155 }  ,fill opacity=1 ] (256.33,10.67) .. controls (256.33,7.81) and (258.65,5.5) .. (261.5,5.5) .. controls (264.35,5.5) and (266.67,7.81) .. (266.67,10.67) .. controls (266.67,13.52) and (264.35,15.83) .. (261.5,15.83) .. controls (258.65,15.83) and (256.33,13.52) .. (256.33,10.67) -- cycle ;
				\draw    (229.67,10.83) -- (256.33,10.67) ;
				\draw [shift={(247.8,10.72)}, rotate = 179.64] [color={rgb, 255:red, 0; green, 0; blue, 0 }  ][line width=0.75]    (8.74,-3.92) .. controls (5.56,-1.84) and (2.65,-0.53) .. (0,0) .. controls (2.65,0.53) and (5.56,1.84) .. (8.74,3.92)   ;
				\draw  [fill={rgb, 255:red, 155; green, 155; blue, 155 }  ,fill opacity=1 ] (293,10.67) .. controls (293,7.81) and (295.31,5.5) .. (298.17,5.5) .. controls (301.02,5.5) and (303.33,7.81) .. (303.33,10.67) .. controls (303.33,13.52) and (301.02,15.83) .. (298.17,15.83) .. controls (295.31,15.83) and (293,13.52) .. (293,10.67) -- cycle ;
				\draw    (266.33,10.83) -- (293,10.67) ;
				\draw [shift={(284.47,10.72)}, rotate = 179.64] [color={rgb, 255:red, 0; green, 0; blue, 0 }  ][line width=0.75]    (8.74,-3.92) .. controls (5.56,-1.84) and (2.65,-0.53) .. (0,0) .. controls (2.65,0.53) and (5.56,1.84) .. (8.74,3.92)   ;
				\draw  [fill={rgb, 255:red, 155; green, 155; blue, 155 }  ,fill opacity=1 ] (330,10.33) .. controls (330,7.48) and (332.31,5.17) .. (335.17,5.17) .. controls (338.02,5.17) and (340.33,7.48) .. (340.33,10.33) .. controls (340.33,13.19) and (338.02,15.5) .. (335.17,15.5) .. controls (332.31,15.5) and (330,13.19) .. (330,10.33) -- cycle ;
				\draw    (303.33,10.5) -- (330,10.33) ;
				\draw [shift={(321.47,10.39)}, rotate = 179.64] [color={rgb, 255:red, 0; green, 0; blue, 0 }  ][line width=0.75]    (8.74,-3.92) .. controls (5.56,-1.84) and (2.65,-0.53) .. (0,0) .. controls (2.65,0.53) and (5.56,1.84) .. (8.74,3.92)   ;
				\draw  [fill={rgb, 255:red, 155; green, 155; blue, 155 }  ,fill opacity=1 ] (367.67,10.67) .. controls (367.67,7.81) and (369.98,5.5) .. (372.83,5.5) .. controls (375.69,5.5) and (378,7.81) .. (378,10.67) .. controls (378,13.52) and (375.69,15.83) .. (372.83,15.83) .. controls (369.98,15.83) and (367.67,13.52) .. (367.67,10.67) -- cycle ;
				\draw    (341,10.83) -- (367.67,10.67) ;
				\draw [shift={(359.13,10.72)}, rotate = 179.64] [color={rgb, 255:red, 0; green, 0; blue, 0 }  ][line width=0.75]    (8.74,-3.92) .. controls (5.56,-1.84) and (2.65,-0.53) .. (0,0) .. controls (2.65,0.53) and (5.56,1.84) .. (8.74,3.92)   ;
				\draw  [dash pattern={on 4.5pt off 4.5pt}]  (378,10.67) .. controls (404.33,12) and (406.33,31.33) .. (378.33,31.33) .. controls (350.33,31.33) and (249.67,29.33) .. (220.33,30) .. controls (191,30.67) and (193,9.33) .. (219.33,10.83) ;

				\draw  [color={rgb, 255:red, 0; green, 0; blue, 0 }  ,draw opacity=1 ][fill={rgb, 255:red, 255; green, 255; blue, 255 }  ,fill opacity=1 ]  (304.26,22.77) -- (536.41,22.77) -- (536.41,62.46) -- (304.26,62.46) -- cycle  ;
				\draw (307.4,27.17) node [anchor=north west][inner sep=0.75pt]  [font=\small,rotate=-0.0]  {$G_{\operatorname{PBC}}( z,\overline{z}) =\int _{-\pi }^{\pi }\frac{\mathrm{d} k}{2\pi }\frac{1}{z-e^{ik}}$};

\end{tikzpicture}%
}

\begin{document}

\title{Universal aspects of bulk density of states in non-Hermitian lattices} 

\author{Mykhailo Pavliuk\,\orcidlink{0009-0005-7729-4013}}
\email{mykhailo.pavliuk@uzh.ch}
\affiliation{Department of Physics, University of Zurich, Winterthurerstrasse 190, 8057 Zurich, Switzerland}

\author{Askar Iliasov\,\orcidlink{0000-0003-2409-7292}}
\affiliation{Department of Physics, University of Zurich, Winterthurerstrasse 190, 8057 Zurich, Switzerland}

\author{Emil J. Bergholtz\,\orcidlink{0000-0002-9739-2930}}
\affiliation{Department of Physics, Stockholm University, SE-106 91 Stockholm, Sweden}

\author{Tom\'{a}\v{s} Bzdu\v{s}ek\,\orcidlink{0000-0001-6904-5264}}
\affiliation{Department of Physics, University of Zurich, Winterthurerstrasse 190, 8057 Zurich, Switzerland}

\begin{abstract}
    Non-Hermitian lattice Hamiltonians generally exhibit strong boundary sensitivity, with periodic  and open boundary conditions producing distinct density of states (DOS) in the complex-energy plane. 
	This has led to the view that extended non-Hermitian systems lack a unique bulk DOS, with different prescriptions representing inequivalent bulk physics.
	Here, we show that this apparent ambiguity is largely illusory. 
	For any finite-range tight-binding Hamiltonian, we establish a universal bulk structure: all DOS definitions arising as thermodynamic limits of finite systems share identical multipole moments and generate identical bulk dynamics at finite times and for observables measured far from boundaries.
	This universality is intimately tied to the thermodynamic Green's functions, which we show to be independent of the boundary condition for large enough complex frequencies.
	Among all equivalent descriptions, we identify the Brown measure---obtained via Hermitization and resolvent analysis---as a canonical and convenient representative of the bulk DOS, defined directly from the infinite-volume Hamiltonian. 
	We further show that point-gap topology imposes additional universal constraints: boundary-dependent Green's functions are forced to coincide throughout topologically trivial point gaps. This, in particular, provides a systematic criterion, valid in arbitrary dimension, for determining where and how eigenvalues of different boundary truncations can accumulate in the complex plane, and precisely delineates the regime in which the DOS ambiguity retains physical significance. 
\end{abstract}

\maketitle

\section{Introduction}

Non-Hermitian Hamiltonians, which in the past decade found applications ranging from classical wave mechanics to open quantum systems~\cite{Ashida:2020,Bergholtz:2021,Song:2019,Kozii:2024,Moiseyev:2011} and were demonstrated in a broad range of experimental setups~\cite{Ruter:2010,Gao:2015,Brandenbourger:2019,photonicNHBBC,Weidemann:2020,Helbig:2020,Liang:2022,Delplace:2017}, present a sharp departure from the familiar Hermitian paradigm: eigenvalues become complex~\cite{Bender:1998,ElGanainy:2018}, left and right eigenvectors differ~\cite{Brody:2014}, and boundary conditions can dramatically change spectra even in the thermodynamic limit~\cite{Hatano:1996,Lee:2016,Yokomizo:2019,Wang:2024,Yang2025,Jezequel2026,Yao:2018,Kunst:2018}. 
The non-Hermitian skin effect~\cite{Yao:2018,Kunst:2018,Kawabata:2020}, in particular, leads to a macroscopic rearrangement of bulk eigenstates under open boundary conditions (OBC), producing spectra that differ strikingly from those under periodic boundary conditions (PBC).
These phenomena have motivated numerous definitions of ``density of states'' (DOS) for non-Hermitian systems, including the PBC DOS obtained from the Bloch states, and the OBC DOS obtained from real-space diagonalization~\cite{Okuma:2020}. 
Non-Hermitian skin effects have been related to nontrivial band topology with respect to ``point gaps'' in the complex energy plane~\cite{Okuma:2020,Zhang:2020,Monkman_2025, Gong:2018}, which differs fundamentally from the ``line-gap'' topology of Hermitian Hamiltonians, and that has been extensively explored~\cite{Gong:2018,Kawabata:2019,Zhou:2019,Kawabata:2019b,Wojcik:2020,Wang:2021,Lieu:2018}.

Because the particular definitions of DOS often yield distinct distributions of eigenvalues in the complex plane, it has been widely accepted that no unique bulk DOS exists for non-Hermitian systems, and that the DOS is fundamentally boundary-condition-dependent. 
With DOS being connected to the Green's function (GF)~\cite{Zirnstein2021,Borgnia2020} and, thus, to the time evolution operator~\cite{abrikosov1964methods}, this notion stands in tension with the physical expectation that bulk dynamics---such as wave packet evolution far from boundaries---should be insensitive to boundary conditions when systems are large. 

In this work, we show that all plausible definitions of non-Hermitian DOS in finite-range tight-binding Hamiltonians are physically equivalent in the bulk. 
In particular, although PBC, OBC, and other constructions produce distinct finite-size spectra, we clarify that Green's functions of these different approximating approaches may be different only inside a bounded region of the complex frequency plane. 
In the thermodynamic limit, these Green's functions coincide for all large enough complex frequencies and therefore possess identical multipole moments~\cite{Cheng_2024}. For instance, OBC DOS can be reconstructed directly from its PBC counterpart whenever the support of the limiting OBC spectrum is known and forms a set of codimension at least one with connected complement.
	
In the time domain, this guarantees that the evolution of any bulk wave packet is insensitive to the way the thermodynamic limit was taken. 
In other words, we show that while being invaluable for the inspection of the non-Hermitian skin effect, different approaches to DOS of extended systems obtained by different limiting procedures have their members operationally indistinguishable at the level of asymptotic bulk observables. 

A natural representative of this class of equivalent DOS is provided by the Brown measure, obtained from the Hermitized resolvent via Girko’s construction~\cite{Girko:1985, Feinberg:1997}. Unlike DOS defined through specific boundary conditions, the Brown measure is: (\emph{i}) intrinsic to the infinite-volume Hamiltonian operator, (\emph{ii}) intimately connected to its spectrum, and (\emph{iii}) independent of how the thermodynamic limit is taken. Approximations to the Brown measure can, in principle, be controllably extracted from non-Hermitian flakes with any boundary conditions. Owing to these special aspects, we argue that the Brown measure constitutes a canonical spectral measure for the non-Hermitian lattice systems.
	
Moreover, the topology of the Hermitized Hamiltonian can be used to determine where boundary-condition-dependent Green's functions may differ, thereby constraining where and how eigenvalues of approximating systems can accumulate. We show that, for arbitrary boundary conditions, such differences can occur only on the spectrum of the infinite operator or within its nontrivial point gaps. They must coincide on all trivial point gaps, including the point gap containing the point at infinity.
Furthermore, since Green's functions are related to the corresponding spectral densities through the two-dimensional Gauss law in the complex plane (see Sec.~\ref{sec:electrostatic_and_GF}), it follows that limiting eigenvalue densities can be supported only on the spectrum of the infinite operator and inside its nontrivial point gaps~\cite{Okuma:2020}. 
Notably, although topologically nontrivial line gaps may generate boundary states accumulating inside otherwise trivial point gaps, the number of such states is subextensive, i.e., they have vanishing eigenvalue density in the thermodynamic limit.
Consequently, they do not contribute to the limiting global eigenvalue density and do not affect the Green's function characterization. 

Our results allow to recast former discussion of spectral collapse in one-dimensional models with non-Hermitian skin effect~\cite{Okuma:2020} with the language of Green's functions and resolvent analysis. 
Crucially, the presented approach applies without modifications to non-Hermitian lattice Hamiltonians in higher dimensions, generally implying universal features of the model's Green's functions and bulk density of states under arbitrary choice of boundary conditions. 
As shown in the later parts of the manuscript, the same framework also captures subextensive spectral features that are invisible in the limiting spectral densities and therefore lie beyond the conventional extensive non-Hermitian skin effect.

The remainder of this article is organized as follows. 
In Sec.~\ref{sec:electrostatic_and_GF}, we introduce our Green's function terminology and the electrostatic analogy. 
In Sec.~\ref{sec:NH_constraints}, we derive the invariant properties of limiting densities and discuss the resulting DOS redundancy in non-Hermitian systems. Sec.~\ref{sec:ex_HN_model} illustrates these ideas using the Hatano-Nelson model. 
In Sec.~\ref{sec:relation_to_evolution}, we trace the origin of this invariance to the invariance of bulk dynamics. 
In Sec.~\ref{sec:brown_measure}, we introduce Hermitization and the Brown measure. 
Then, in Sec.~\ref{sec:limsup}, we clarify the role of the point-gap topology in constraining which thermodynamic limits produce DOS differing from the PBC DOS. 
In Sec.~\ref{sec:energy_accumulation} we elaborate on universal aspects of energy level accumulation beyond the point-gap topology. 
Finally, Sec.~\ref{sec:sym-protected} extends this discussion to systems with inherent non-Hermitian symmetries.

\section{Equivalence of various definitions \texorpdfstring{\nopagebreak\\}{} of bulk DOS in non-Hermitian systems}

\subsection{Green's function, spectral density, \texorpdfstring{\nopagebreak\\}{} and their relation in non-Hermitian systems} \label{sec:electrostatic_and_GF}

Assume a general non-Hermitian lattice Hamiltonian $\hat{H}_{\Omega_N}$ corresponding to some finite flake of the size $N$ of the infinite lattice; $\Omega$ represents (symbolically) how exactly this flake was obtained and boundary conditions. 
Define its resolvent as:
\begin{equation}\label{eqn:resolvent-def}
    \hat{G}_{\Omega_N}(z)=\frac{1}{z-\hat{H}_{\Omega_N}},\quad z\in\mathbb C\backslash\operatorname{spec}(\hat{H}_{\Omega_N}).
\end{equation}
This quantity allows for systematic access to eigenvalues of the matrix $\hat{H}_{\Omega_N}$. 
To do so, one focuses on the scalar trace of the resolvent,
\begin{equation} \label{eq:resolvent-trace}
    G_{\Omega_N}(z) \equiv \frac{1}{N}\Tr \frac{1}{z-\hat{H}_{\Omega_N}}=\frac{1}{N}\sum_{i=1}^N\frac{1}{z-z_i},
\end{equation}
which we will also refer to simply as the Green’s function. 
Note that this should be distinguished from the local Green’s function, introduced later in the text. 
In the last expression, $z_i$ denote the zeros of the characteristic polynomial. 

Thus, in finite-size systems, eigenvalues of the lattice Hamiltonian appear as the poles of the Green's function.
In Eq.~(\ref{eq:resolvent-trace}), we implicitly assume that the sum counts each eigenvalue according to its algebraic multiplicity. 

Global normalized DOS can be systematically obtained from $G_{\Omega_N}(z)$ by applying the anti-holomorphic derivative to to Eq.~(\ref{eq:resolvent-trace}), and using the distributional identity $\frac{\partial }{\partial \bar z}\frac{1}{z-z_0}=\pi \delta(z-z_0)$:
\begin{equation}\label{GF_spec_formula}
    \frac{\partial }{\partial \bar z} G_{\Omega_N}(z) = \frac{1}{N}\sum_{i=1}^N \pi \delta(z-z_i)\equiv \pi \rho_{\Omega_N}(z)
\end{equation}
with $\rho(z)$ normalized to satisfy the standard sum rule $\int_\mathbb{C} \mathrm{d}^2z \, \rho_{\Omega_N}(z) = 1$. 

One can now take the thermodynamic limit (TL) of Eq.~(\ref{GF_spec_formula}). 
If the Green's function and the spectral density converge (in the distributional sense), relation (\ref{GF_spec_formula}) continues to hold for the limiting objects. 
Moreover, the limiting Green's function (and spectral density) may appear to be represented by continuous (but not analytic) functions in the whole $\mathbb{C}$. 
To emphasize this aspect, we write the limiting form of both of them as explicitly dependent on both $z$ and $\bar z$:
\begin{equation}\label{eq:gauss_law_limit}
        \frac{\partial }{\partial \bar z} G_{\Omega}(z,\bar z) = \pi \rho_{\Omega}(z, \bar z)
\end{equation}
where we keep explicit dependence on $\Omega$ to remember that the limiting DOS (as well as the Green's function) depends on how the thermodynamic limit 
was taken.
Equivalently, the last relation can be rewritten in the form:
\begin{equation}\label{eq:disp_rel_form}
         G_{\Omega}(z,\bar z) = \int_\mathbb{C}  \mathrm d^2 \omega \ \frac{\rho_{\Omega}(\omega, \bar \omega)}{z-\omega}.
\end{equation}
These relations between the spectral density and the Green’s function may at first appear unfamiliar, as they are not written in the standard form used for Hermitian systems [in particular, Eq.~(\ref{eq:gauss_law_limit})]. There is, however, a useful analogy that allows one to clarify potential misconceptions in a straightforward way.

To make this analogy explicit, introduce real variables $x,y$ such that $z=x+iy$, define $ \boldsymbol\nabla=(\frac{\partial}{\partial x}, \frac{\partial}{\partial y})$ and $\boldsymbol{G}(x,y)=(\Re{G_\Omega(z)},-\Im{G_\Omega(z)})$. 
Equation~(\ref{eq:gauss_law_limit}) then takes the form of the two-dimensional Gauss law,
\begin{equation}
    \boldsymbol{\nabla}\cdot\boldsymbol{G}_\Omega=\pi \rho_\Omega.
\end{equation}
This shows that the Green’s function and the corresponding spectral density stand in the same relation as the electric field and the charge density that generates it in two-dimensional (2D) electrostatics.

In direct analogy with electrostatics, the limiting spectral distribution may be surface-supported, volume-supported (as anticipated in higher-dimensional non-Hermitian models), or mixed.\footnote{Here we understand `surfaces' as 1D while `volumes' as 2D subspaces of the of the complex plane.}
For example, in Hermitian systems, the limiting distribution is of surface type and confined to the real axis. Gauss’ law then implies the usual Maxwell interface conditions, which in this context yield the standard discontinuity (“jump”) of the Green’s function across the real line: $G(\omega+i\delta)-G(\omega-i\delta)=2\pi \rho(\omega)$. This also clarifies how Eqs.~(\ref{eq:gauss_law_limit}) and~(\ref{eq:disp_rel_form}) reduce to the textbook form for Hermitian systems~\cite{abrikosov1964methods}.

Before proceeding, we further introduce the local Green’s function,\footnote{For lattices with multiple orbitals per unit cell, the subsequent analysis can be carried out by replacing the local expectation value $\bra{i}\,\cdot\,\ket{i}$ 
with the normalized trace over the unit-cell degrees of freedom,
\[
\bra{i}\,\cdot\,\ket{i}\;\rightarrow\;\frac{1}{M}\sum_{\alpha=1}^M \bra{i,\alpha}\,\cdot\,\ket{i,\alpha}.
\]
}
\begin{equation}\label{eq:local_GF_def}
G_{\Omega_N}(i,i;z)=\bra{i}\frac{1}{z-\hat H_{\Omega_N}}\ket{i}.
\end{equation}
In Hermitian systems, the local Green’s function, Eq.~(\ref{eq:local_GF_def}), and the normalized trace of the resolvent, Eq.~(\ref{eq:resolvent-trace}), are expected to coincide in the thermodynamic limit, provided $\Im z \neq 0$. Since the local Green’s function is often easier to compute, it is natural to shift attention from the global Green’s function to its local counterpart. However, such a relation between these two objects in non-Hermitian context is missing, due to boundary sensitivity.

From a physical perspective, the local Green’s function is usually the more relevant quantity, since it is directly related to transition amplitudes and local response properties. At the same time, defining a non-Hermitian analogue of the local density of states for finite flakes is problematic. Indeed, in a generic non-Hermitian model, $\bar \partial G_{\Omega_N}(i,i;z)$ is not given by a simple weighted sum of delta functions, in contrast to $\bar \partial G_{\Omega_N}(z)$ as implied by Eq.~(\ref{GF_spec_formula}). Instead, $\bar \partial G_{\Omega_N}(i,i;z)$ may contain derivatives of delta functions,  %
and therefore need not admit an interpretation as a genuine density for any finite flake $\Omega_N$. As a result, for a given sequence of flakes ${\hat{H}_{\Omega_N}}$, the corresponding limiting local spectral \textit{density} may fail to exist. %

At first sight, this leads to a somewhat unsatisfactory situation: the spectral density is naturally defined through the trace of the resolvent, whereas the local Green’s function is the quantity of more direct physical relevance. The key point, however, is that these two objects can be shown to coincide for sufficiently large $|z|$ [see~Sec.~\ref{sec:limsup}]. In the electrostatic language, this means that the far fields of $G_{\Omega_N}(z)$ and $G_{\Omega_N}(i,i;z)$ coincide in the thermodynamic limit. This observation will allow us to infer properties of $G_{\Omega}(z,\bar z)$, and hence of $\rho_{\Omega}(z,\bar z)$, from the physical properties of the simpler and more accessible local Green’s function $G_{\Omega}(i,i;z)$.

\subsection{Constraints on possible bulk DOS \texorpdfstring{\nopagebreak\\}{} of extended non-Hermitian systems} \label{sec:NH_constraints}

It is a well-known observation that, contrary to the Hermitian case, non-Hermitian systems are very sensitive 
to the precise form of the boundary conditions to which the system is  
subjected under the procedure of taking the thermodynamic limit. 
Specifically, eigenvalues accumulate differently in the complex plane, producing different limiting spectral distributions.

Since the spectral density and the trace of the resolvent are related through Eq.~(\ref{eq:disp_rel_form}), the latter also appears to depend on the precise manner in which the thermodynamic limit is taken.

This conclusion may nevertheless seem surprising, if one expects the trace of the resolvent, Eq.~(\ref{eq:resolvent-trace}), and the local Green's function, Eq.~(\ref{eq:local_GF_def}), to be connected in the thermodynamic limit, as they certainly are in the Hermitian case.
The local Green's function, or more generally,
\[G_{\Omega_N}(i,j;z)=\bra{i}(z-\hat H_{\Omega_N})^{-1}\ket{j},\] 
is closely related to the evolution operator (see Sec.~\ref{sec:relation_to_evolution}) and therefore encodes, in particular, the evolution of states localized in the bulk. One would expect such bulk dynamics %
to be insensitive to boundary conditions. If so, this insensitivity should be reflected in the limiting local GF and, consequently, in the trace of the resolvent, if any connection is expected between the two.

If this expectation is taken seriously, it implies an implicit constraint on the set of spectral densities $\rho_\Omega(z,\bar z)$ that can arise in the thermodynamic limit of a given lattice model. In what follows, we derive this constraint. In Sec.~\ref{sec:relation_to_evolution}, we explain its physical origin and relate it to the insensitivity of bulk dynamics to boundary conditions.

Consider the sequence of approximating Hamiltonians, $\{\hat H_{\Omega_N}\}$, which is constructed to approximate an infinite lattice Hamiltonian $\hat{H}$ (we restrict ourselves to the case where $\hat{H}$ is a tight-binding Hamiltonian with finite-range hoppings, so that $\hat{H}$ is bounded). Formally, one demands $\hat H_{\Omega_N}\rightarrow \hat H$ in the appropriate sense; namely, we demand $\hat H_{\Omega_N}$ to converge strongly\footnote{\label{foot:strong-convergence}Sequence of (bounded) operators $\hat H_{\Omega_N}$ converges to $\hat{H}$ strongly if for any (normalizable) state $\ket{\psi}$, $\norm{\hat H_{\Omega_N} \ket{\psi} - \hat H \ket \psi} \rightarrow 0$. Note that when phrasing this statement, we assume that each approximation $\hat{H}_{\Omega_N}$ is acting in the infinite-dimensional Hilbert space of $\hat{H}$; namely, $\hat{H}_{\Omega_N}$ acts non-trivially only on the spatially finite domain $\Omega_{N}$ and vanishes outside it. Throughout the manuscript, unless stated otherwise, convergence of a sequence of operators is understood as strong convergence in the sense specified above.} 
 to $\hat{H}$. 
This is a physically adequate requirement, satisfied for most standard choices of the boundary conditions.

The resolvent $(z-\hat{H})^{-1}$ of such a bounded operator $\hat H$ exists as a well-defined operator outside of the spectrum\footnote{The spectrum of the bounded operator \(\hat H\) is understood as the subset of the complex plane consisting of all \(z\) for which \(z-\hat H\) is not bijective, cf.~Eq.~(\ref{eqn:resolvent-def}).}
of the operator $\hat{H}$. In particular, for $z>||\hat{H}||$ (where $||\hat{H}||$ denotes the norm of the operator\footnote{Here, the norm of the operator is understood in the standard operator-norm sense, i.e.,
\[
\|\hat H\|=\sup\{\|\hat H\psi\|:\|\psi\|=1\}.
\]
}) the resolvent is in fact represented by the convergent Neumann series~\cite{reed2012methods}: 
\begin{equation}\label{eq:neumann_series}
    \frac{1}{z-\hat H}=\frac{1}{z}\cdot\sum_{n=0}^\infty\left(\frac{\hat{H}}{z}\right)^n.
\end{equation}
Specifically, the local resolvent of the infinite-lattice Hamiltonian, 
\begin{equation}
	G(z)=\bra{i}(z-\hat H)^{-1}\ket{i},
\end{equation}
is given for these $z$ by absolutely convergent power series on $1/z$:
\begin{equation}\label{eq:multipole_expansion}
    G(z)=\sum_{n=1}^\infty a_n  \left(\frac{1}{z}\right)^n
\end{equation}
with $a_1=1$ and $a_{n} =\bra{i}\hat{H}^{n-1}\ket{i}$ for $n>1$. Note that, analogous to the Hermitian case, $\bra{i}\hat{H}^{n-1}\ket{i}$ counts the number of self-returning walks, weighted by corresponding hopping amplitudes.

Using the Neumann series in Eq.~(\ref{eq:neumann_series}), one can prove the following fact (see also Sec.~\ref{sec:limsup}): if $\hat H_{\Omega_N}$ converges strongly to $\hat H$, then it is guaranteed that there exists a finite $M\geq||\hat{H}||$ such that strong resolvent convergence (SRC) is implied for all $|z|>M$. In fact, $M=\limsup_N\norm{\hat H_{\Omega_N}}$.

In other words, no matter which boundary conditions are chosen, for all $|z|>M$ one has $(z-\hat{H}_{\Omega_N})^{-1}\rightarrow (z-\hat{H})^{-1}$. In particular, for such $z$, the limiting local Green's function coincides with $\bra{i}(z-\hat H)^{-1}\ket{i}$:
\begin{equation}\nonumber
    G_{\Omega_N}(i,i;z)\rightarrow G(z).
\end{equation}
However, as already foreshadowed at the end of Sec.~\ref{sec:electrostatic_and_GF} (and to be clarified in Sec.~\ref{sec:limsup}), the limiting local Green’s function in the regime $|z|>M$ coincides with the limiting trace of the resolvent; therefore also:
\begin{equation}\label{GF_far-field_convergence}
    G_{\Omega_N}(z)\rightarrow G(z),\ \text{for } z \text{ satisfying } |z|>M.
\end{equation}

Note that inside the region $|z|<M$, $G_{\Omega_N}(z)$ need not to converge to $G(z)$, even for those $z$ for which $(z-\hat H)^{-1}$ is defined. In particular, for such $z$, one does not in general expect two different thermodynamic limits, $\Omega$ and $\tilde\Omega$, to share the same limiting GF; that is, generally, 
\begin{equation}\nonumber
    G_{\overset{\ }{\Omega}}(z,\bar z)\neq G_{\tilde \Omega}(z,\bar z),\quad|z|<M.
\end{equation}
On the other hand, in the region $|z|>M$, the functions $G_{\overset{\ }{\Omega}}(z,\bar z)$ and $ G_{\tilde \Omega}(z,\bar z)$ coincide and are both equal to the analytic function $G(z)$, given by Eq.~\eqref{eq:multipole_expansion}.

In the electrostatic parlance of Sec.~\ref{sec:electrostatic_and_GF}, this observation can be reformulated in an especially straightforward manner. 
While different thermodynamic limits may produce different limiting Green's functions $G_\Omega(z)$ (the ``electric fields''), any pair of limiting GFs should coincide for all high enough complex frequencies. 
Motivated by the electrostatic analogy, we say in such cases that both GFs share the same far-field configuration (even though their near-field configurations may differ). Alternatively, any pair of limiting spectral densities (the ``charge densities'') corresponding to different ways of taking TL of the same physical system is forced to produce the same configuration of GFs in the far-field region.

This observation is completely analogous to usual electrostatics, where it is well-known that the far field doesn't uniquely determine the charge configuration (e.g., both a point charge $Q$ and a uniformly spherically distributed charge $Q$ produce the same far field). 
Yet, different charge configurations producing the same far fields necessarily share identical multipole moments. Analogously, we can say that while different TLs may produce different limiting spectral densities $\rho_\Omega(z,\bar z)$, all of their 2D multipole moments
\begin{equation}\label{eqn:spectral-moments}
    q_m=\int_{\mathbb C} \mathrm d^2z\ z^m\rho_\Omega(z,\bar z)
\end{equation}
are the same.

Therefore, multipole moments of the spectral density $q_m$ are exact invariants and are not sensitive to the way the TL was obtained. 
Using the representation in Eq.~(\ref{eq:multipole_expansion}) together with the complex version of Green’s theorem applied to the Gauss law, Eq.~(\ref{eq:gauss_law_limit}), we obtain the following simple expression for the multipole moments:
\begin{equation}
    q_m = a_{m+1} =\bra{i}\hat{H}^{m}\ket{i},
\end{equation}
which, when placed in the Hermitian context, is the familiar relation 
connecting, for example, the number of closed walks with the moments of DOS in the nearest-neighbor models.\footnote{In particular, this reduces to an equivalent but more involved integral over the Brillouin zone~\cite{Cheng_2024}.}

	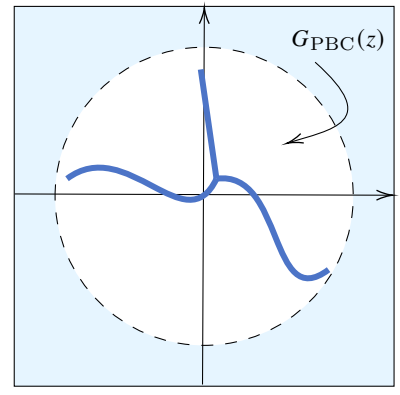
\begin{figure}
		\centering
		\begin{tikzpicture}[x=0.75pt,y=0.75pt,yscale=-1,xscale=1,scale=0.75]
\draw  [color={rgb, 255:red, 0; green, 0; blue, 0 }  ,draw opacity=1 ][fill={rgb, 255:red, 230; green, 245; blue, 255 }  ,fill opacity=1 ] (187.67,15.17) -- (441.83,15.17) -- (441.83,269.33) -- (187.67,269.33) -- cycle ;
\draw  [fill={rgb, 255:red, 255; green, 255; blue, 255 }  ,fill opacity=1 ][dash pattern={on 4.5pt off 4.5pt}] (215,142.25) .. controls (215,87.16) and (259.66,42.5) .. (314.75,42.5) .. controls (369.84,42.5) and (414.5,87.16) .. (414.5,142.25) .. controls (414.5,197.34) and (369.84,242) .. (314.75,242) .. controls (259.66,242) and (215,197.34) .. (215,142.25) -- cycle ;
\draw    (313.66,268.19) -- (314.88,17.84) ;
\draw [shift={(314.89,15.84)}, rotate = 90.28] [color={rgb, 255:red, 0; green, 0; blue, 0 }  ][line width=0.5]    (10.93,-3.29) .. controls (6.95,-1.4) and (3.31,-0.3) .. (0,0) .. controls (3.31,0.3) and (6.95,1.4) .. (10.93,3.29)   ;
\draw    (188,140.67) -- (439,141.53) ;
\draw [shift={(441,141.53)}, rotate = 180.2] [color={rgb, 255:red, 0; green, 0; blue, 0 }  ][line width=0.5]    (10.93,-3.29) .. controls (6.95,-1.4) and (3.31,-0.3) .. (0,0) .. controls (3.31,0.3) and (6.95,1.4) .. (10.93,3.29)   ;
\draw [color={rgb, 255:red, 76; green, 116; blue, 199 }  ,draw opacity=1 ][line width=2.25]    (223,130.5) .. controls (263,100.5) and (302.5,173.84) .. (323,130.5) ;
\draw [color={rgb, 255:red, 76; green, 116; blue, 199 }  ,draw opacity=1 ][line width=2.25]    (323,130.5) .. controls (368.5,124.34) and (358,221.34) .. (398,191.34) ;
\draw [color={rgb, 255:red, 76; green, 116; blue, 199 }  ,draw opacity=1 ][line width=2.25]    (323,130.5) -- (312.5,57.34) ;
\draw    (389,52.5) .. controls (434.57,93.02) and (398.99,97.67) .. (371.66,106.31) ;
\draw [shift={(370,106.84)}, rotate = 341.88] [color={rgb, 255:red, 0; green, 0; blue, 0 }  ][line width=0.5]    (10.93,-3.29) .. controls (6.95,-1.4) and (3.31,-0.3) .. (0,0) .. controls (3.31,0.3) and (6.95,1.4) .. (10.93,3.29)   ;

\draw (371.5,28.4) node [anchor=north west][inner sep=0.75pt]    {$G_{\operatorname{PBC}}( z)$};

\end{tikzpicture}

		\caption{Reconstruction of the OBC spectrum from the PBC spectrum. The horizontal (vertical) axis correspond to the real (imaginary) part of $z$.
        Assuming the limiting OBC spectrum is confined to a known codimension-1 subset of  $\mathbb{C}$ (thick blue curve), the corresponding OBC Green’s function can be obtained by taking the PBC Green’s function in the far-field region (the exterior of the dashed disk) and analytically continuing it into the small-$|z|$ region. The OBC density of states is then recovered from the boundary values of the Green’s function.}
        \label{fig:OBC_reconstruction}
	\end{figure}
    
Hermitian systems are quite special here, as their Green's functions are forced to be analytic functions on $\mathbb{C}\backslash\operatorname{spec}(\hat H)$, i.e., on $\mathbb{C}$ except at most some bounded interval of $\mathbb R$. 
Any sequence of approximating Green's functions $G_{\Omega_N}(z)$ can only converge to the analytic function. 
However, relation~(\ref{GF_far-field_convergence}) along with the uniqueness of the analytic continuation fixes this limiting function to coincide with $G(z)=\bra{i}(z-\hat H)^{-1}\ket{i}$ for \emph{all} $z\in \mathbb{C}\backslash\mathbb R$, as the analytic content of this limiting function is fully encoded in its far-field configuration. 

An alternative way to phrase this is as follows. For a general model with limiting spectral density $\rho_\Omega(z,\bar z)$, the far-field behavior determines only the multipole data, namely the harmonic moments of the two-dimensional density. These are the moments obtained by integrating $\rho_\Omega$ against harmonic polynomials $p(z,\bar z)$ satisfying $\partial\bar\partial p=0$. By contrast, the far field does not determine the full set of \emph{mixed} moments
\begin{equation}
\label{eqn:mixed-moments}
\int d^2 z\, z^m \bar z^n \rho_\Omega(z,\bar z).
\end{equation}
If it did, then a compactly supported density $\rho_\Omega(z,\bar z)$ would be uniquely recoverable from its far-field behavior alone. The Hermitian case is exceptional because the support of the limiting measure is restricted to the real axis. This additional constraint eliminates the freedom in transverse moments, so that the multipole moments completely determine the ordinary one-dimensional moments of the density along $\Re[z]$.

Another important consequence that one can deduce from such an analysis is the following. 
Assume that in some particular TL (say, with OBC), it is guaranteed that the spectra are real (e.g., in models with unbroken $\mathcal{PT}$-symmetry), or at least confined to a known subset in the complex plane with connected complement. 
Suppose further that this subset has codimension at least~1.
Then one can always restore OBC spectra from any other TL spectra (e.g., from the PBC one) by doing analytic continuation of $G(z)$ from the far-field region and restoring spectral density from boundary values of $G(z)$ (see Fig.~\ref{fig:OBC_reconstruction}). 
In particular, if the function cannot be analytically continued to have a cut on the real line, then the autocorrelation function behavior is inherently non-Hermitian and is not expected to show the usual Hermitian dispersive behavior~\cite{Cheng_2024}.

\subsection{Example: Hatano-Nelson model} \label{sec:ex_HN_model}

Both previous points can be easily demonstrated on the example of the Hatano-Nelson (HN) model~\cite{Hatano:1996}: 
\begin{equation}\label{eq:HN_Hamiltonian}
    \hat{H}=\sum_i \left( t_R \ket{i+1}\bra{i} +t_L\ket{i}\bra{i+1} \right).
\end{equation}
For simplicity, first consider the unidirectional limit: $t_R\,{=}\,1$, $t_L\,{=}\,0$. 
If subjected to PBC, the Green's function of the model is well-known and is given by 
\begin{equation}\label{eqn:unidirectional-HN-PBC}
    G_{\mathrm{PBC}}(z,\bar z)=\frac{1}{2\pi}\int_0^{2\pi} \mathrm d k \frac{1}{z-e^{ik}}
\end{equation} 
in the thermodynamic limit. 
Taking $\bar \partial$ of this expression, and using $\bar \partial\frac{1}{z-\omega}=\delta(z-\omega)$ together with the relation (\ref{eq:gauss_law_limit}), one obtains the PBC spectral function $\rho_{\mathrm{PBC}}(z, \bar z)=\frac{1}{2\pi}\int_{0}^{2\pi} \delta(z-e^{ik})$, i.e., a uniformly charged ring of unit radius, in electrostatic parlance. 
Using either a direct computation of the integral or the application of 2D Gauss law for a radially symmetric system, one obtains the Green's function:
\begin{equation} \label{eq:unidirectional_HN_PBC_GF}
    G_{\mathrm{PBC}}(z,\bar z)
    =
    \begin{cases}
        \frac{1}{z}, & |z|>1,\\
        0, & |z|<1
    \end{cases}
\end{equation}

\begin{figure}[t!]
\centering

\begin{minipage}[t]{0.48\columnwidth}
  \centering
  \includegraphics[width=\linewidth]{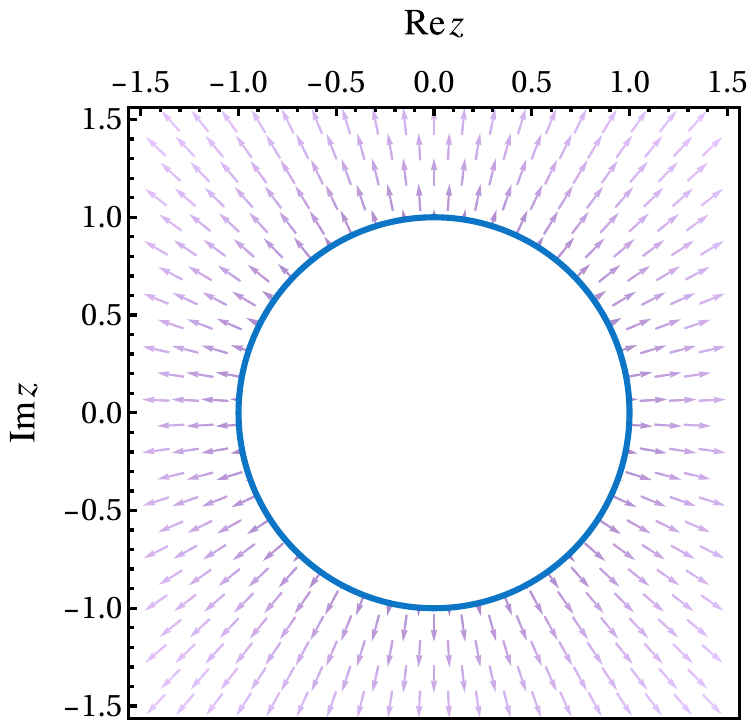}
\end{minipage}
\hfill
\begin{minipage}[t]{0.48\columnwidth}
  \centering
  \includegraphics[width=\linewidth]{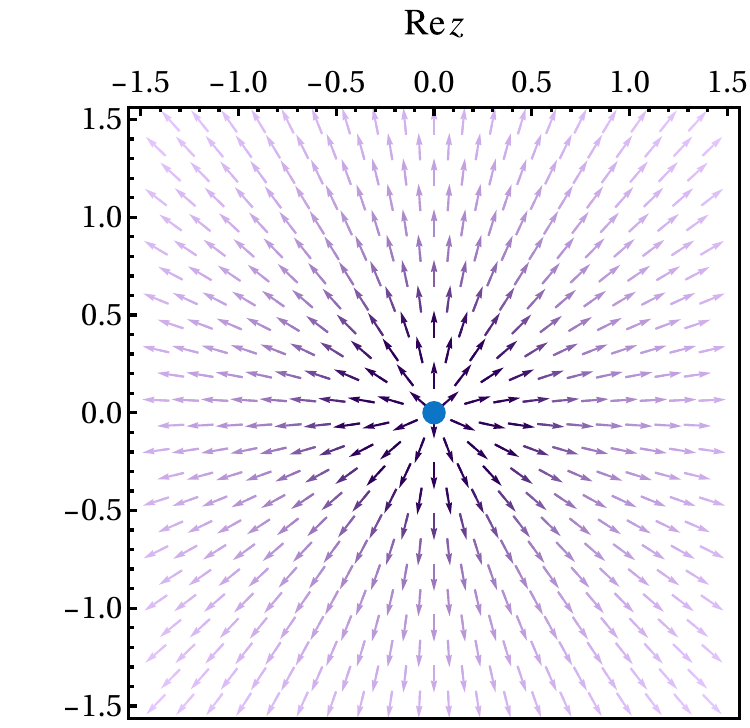}
\end{minipage}

\vspace{-45mm}
\begin{minipage}[t]{0.48\columnwidth}
\subfloat[]{\label{fig:pbc}}\hfill %
  \vspace{-2mm} %
\end{minipage}
\hfill
\begin{minipage}[t]{0.48\columnwidth}
  \subfloat[]{\label{fig:obc}}\hfill %
  \vspace{-2mm}
\end{minipage}

\vspace{35mm}

\begin{minipage}[t]{0.48\columnwidth}
  \centering
  \resizebox{\linewidth}{!}{%
    \PBCtikz
  }
\end{minipage}
\hfill
\begin{minipage}[t]{0.48\columnwidth}
  \centering
  \resizebox{0.88\linewidth}{!}{%
    \OBCtikz
  }
\end{minipage}

\caption{Unidirectional Hatano--Nelson model. (a) PBC: 
$G_{\mathrm{PBC}}(z,\bar z)$ is equal to $1/z$ outside the unit circle, $|z|>1$, and vanishes inside it.
(b) OBC: $G_{\mathrm{OBC}}(z,\bar z)$ is given by $1/z$ everywhere in the complex plane. The corresponding spectral distributions are schematically illustrated by a blue circle and a blue point, respectively. The Green’s functions are represented by the vector fields $\mathbf G(z)=\left(\Re G(z),-\Im G(z)\right)$. The color saturation of the vectors encodes the magnitude of the vector field.}
\label{fig:unidirectional_1D_HN_model}

\end{figure}

On the other hand, the OBC GF of the unidirectional HN chain is a simple Jordan block, and thus: 
\begin{equation}
    G_{\operatorname{OBC}}(z)=\frac{1}{z},\quad |z|>0.
\end{equation}
This is schematically demonstrated in Fig.~\ref{fig:unidirectional_1D_HN_model}. Note how two different GFs share the same far-field configurations while being produced by markedly different ``charge densities''. 
Note also that trying to analytically continue the PBC GF Eq.~(\ref{eq:unidirectional_HN_PBC_GF}) from the far-field region into the region $|z|<1$, one recovers the OBC GF.

A somewhat more involved example of this kind, which may demonstrate all the points much more explicitly, is the bidirectional HN chain. The PBC GF is again given by:
\begin{equation}\label{eq:general_HN_starting_form}
    G_{\mathrm{PBC}}(z,\bar z) = \frac{1}{2\pi}\int_0^{2\pi} \mathrm d k \frac{1}{z-t_R e^{ik}-t_L e^{-ik}}.
\end{equation}
We evaluate this integral in Appendix~\ref{a:evaluation_of_the_PBC_GF}. The final result is given by the following expression: 
\begin{equation}\label{eq:PBC_GF_HN}
 G_{\mathrm{PBC}}(z,\bar z)=
    \begin{cases}
            \frac
    {
        1
    }
    {
        \sqrt{z^2-4t_Rt_L}
    },& |z-c|+|z+c|\geq2a  \\
    0,& \text{otherwise,}\\
    \end{cases}
\end{equation}
where $a=t_R+t_L$ and $c=2\sqrt{t_Rt_L}$.
On the other hand, as is well known, OBC Hatano-Nelson is pseudo-Hermitian, being related by a 
similarity transformation to the Hermitian OBC problem with effective hopping $t=\sqrt{t_R t_L}$.
Its TL GF is therefore that of an infinite Hermitian chain with an effective reciprocal hopping $\sqrt{t_R t_L}$:
\begin{equation}\label{eq:OBC_GF_HN}
        G_{\mathrm{OBC}}(z)=
    \frac
    {
        1
    }
    {
        \sqrt{z^2-4t_Rt_L}
    },\ z\in \mathbb{C}/[-c,c]
\end{equation}
and this holds for all $z$ outside the real line (not only outside the ellipse, as in the PBC case).
One can see again that the far-field configuration of both the PBC and the OBC Green's functions coincide, while being completely different in the near-field region. 
Note that the OBC GF coincides with the analytic continuation of the PBC GF [Eq.~(\ref{eq:PBC_GF_HN})] from the far-field region to the whole complex plane, in accordance with the conclusions stated at the end of Sec.~\ref{sec:NH_constraints}. %

Alternatively, both GFs can be thought of as created by their corresponding spectral distributions. Namely, the PBC one is created by the following ``charge density'':
\begin{equation}
    \rho_{\operatorname{PBC}}(z,\bar z)=\frac{1}{2\pi}\int_{-\pi}^\pi \mathrm{d}k\ \delta(z-\epsilon(k)),
\end{equation}
i.e., by the charged ellipse, with the line charge density per unit arclength 
$$\sigma_{\operatorname{PBC}}(\phi)=\frac{1}{2\pi}\frac{1}{\sqrt{t_R^2+t_L^2-2t_Rt_L \cos{(2\phi)}}},$$
where $\phi$ is the standard elliptic parameter along the ellipse.

The following charge density produces the OBC GF:
\begin{align}\nonumber
    \rho_{\operatorname{OBC}}(z,\bar z)=\frac{1}{\pi}&\frac{1}{\sqrt{4t_Rt_L-(\Re z)^2}}
    \\
    &\quad\times\delta(\Im{z})\Theta(4t_Rt_L-(\Re z)^2),
\end{align}
where $\Theta:\mathbb{R}\to\{0,1\}$ is the Heaviside function. The derived result corresponds to %
a charged rod supported on the real segment between the foci of the same ellipse. Its nonuniform line charge density is given by the first factor in the above expression,
which coincides with the density of states of an infinite Hermitian chain with reciprocal hopping $\sqrt{t_R t_L}$.

\begin{figure}[t!]
\includegraphics[width=1\linewidth]{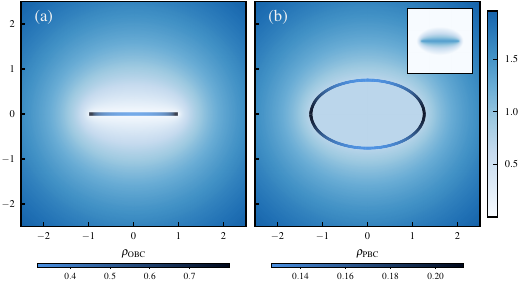}
\caption{Bidirectional Hatano--Nelson model. For convenience, its ``electrostatic potential'' $\Phi(z)$ is shown, with $\Phi(z)$ satisfying $G(z)=\partial\Phi(z)/\partial z$. 
Along with the potential, the corresponding spectral functions are depicted in terms of the line charge density per unit arclength, which serve as the sources of the potential. 
For both the PBC and the OBC systems, we have set $t_R=1,t_L=0.25$. 
(a)~For OBC, the potential is $\Phi(z)=\Re \cosh^{-1}(z/c)$.
(b)~For PBC, the potential is the same outside the elliptic spectral region, while it is constant inside it. The inset figure shows the difference $\Phi_{\mathrm{OBC}}-\Phi_{\mathrm{PBC}}$, which is exactly zero outside the PBC elliptic spectrum.
}
\label{fig:bidirectional_1D_HN_model}
\end{figure}

It turns out that both distributions produce the same far fields and thus share the same multipole moments.
The comparison is illustrated in Fig.~\ref{fig:bidirectional_1D_HN_model}, where we plot the real-valued ``potential'' \footnote{The potential satisfies \(G(z)=\partial \Phi(z)/\partial z\) and is uniquely determined (up to an additive constant) by its \(\log |z|\) behavior at infinity.} $\Phi(z)$ associated with the corresponding Green’s functions, together with schematic representations of the respective spectral distributions.

\subsection{Physical origin of DOS equivalence}\label{sec:relation_to_evolution}

It is instructive to understand the physical origin of this ambiguity and what its physical meaning is. 
While the precise physical content of the eigenenergies and eigenstates is not always immediately transparent, especially in non-Hermitian systems, where their interpretation can be more subtle, the physical content of the resolvent $(z-\hat{H})^{-1}$ is relatively clear, as this object is tightly connected to the evolution operator $e^{-i\hat{H}t}$, which dictates, e.g., how states propagate.

Physically, one anticipates the following: assume that (\emph{i})~some BC are imposed, that (\emph{ii})~the sequence of approximating Hamiltonians is given (say, specified by $\hat{H}_{\Omega_N}$), and that (\emph{iii})~the TL of given sequence reproduces the ideal infinite lattice system described by the bounded Hamiltonian $\hat{H}$ (in the sense that $\hat{H}_{\Omega_N}\rightarrow\hat{H}$ strongly). 
Then, when trying to propagate an arbitrary normalizable state that is sufficiently far  from the boundaries, one expects that the evolution of such a state under
the approximating Hamiltonian should be almost indistinguishable from an evolution under the ideal limiting  Hamiltonian.
Strictly speaking, one expects that if $\hat{H}_{\Omega_N}\rightarrow\hat{H}$, then $e^{-i\hat{H}_{\Omega_N}t}\rightarrow e^{-i \hat H t}$ as well for each fixed $t$; this, in fact, is true for bounded operators, which confirms the physical intuition.

However, as was already mentioned, the evolution operator is tightly connected to the resolvent operator. Therefore, the fact that the evolution operators of the approximating systems converge to the evolution operator of the limiting one should impose certain constraints on the convergence behavior of the resolvent operators. In fact, such a constraint is, roughly speaking, the requirement of the same limiting far-field configuration of resolvents.
This can be most simply demonstrated by using the following relation between the resolvent and the evolution operator, which holds for an arbitrary bounded operator $\hat{H}$~\cite{reed2012methods}: 
\begin{equation}\label{eq:dunford_integral}
    e^{-i\hat{H}t}=\frac{1}{2\pi i}\oint_{\Gamma}\frac{e^{-izt}}{z-\hat{H}}\mathrm{d}z,
\end{equation}
where $\Gamma$ is a closed counterclockwise contour that encircles the spectrum of $\hat{H}$ once. 
In particular,
\begin{equation}\label{eq:dunford_integral_matrix_element}
    \bra{i}e^{-i\hat{H}t}\ket{i}
    =\frac{1}{2\pi i}\oint_{\Gamma}e^{-izt} G(z)\mathrm{d}z,
\end{equation}
i.e., the return amplitude is a certain contour integral of the Green's function. An analogous formula also holds for the off-diagonal elements.

One can now see that the coincidence of the limiting resolvents in the far-field region essentially ensures the consistency of representation~(\ref{eq:dunford_integral}) with the convergence $e^{-i\hat{H}_{\Omega_N}t}\rightarrow e^{-i\hat{H}t}$.
Assume, e.g., that the  spectrum of both $\hat{H}$ and the approximating sequence is contained in some ball of radius $M$. Choose an arbitrary contour $\Gamma$ outside of this ball; as already mentioned, in this region $(z-\hat{H}_{\Omega_N})^{-1}\rightarrow(z-\hat{H})^{-1}$ (uniformly on the contour). 
Therefore, one also has
$$\oint_{\Gamma}\frac{e^{-izt}}{z-\hat{H}_{\Omega_N}}\mathrm{d}z\rightarrow\oint_{\Gamma}\frac{e^{-izt}}{z-\hat{H}}\mathrm{d}z$$
for all such contours, confirming that $e^{-i\hat{H}_{\Omega_N}t}\rightarrow e^{-i\hat{H}t}$.

Essentially, the argument above shows that it is the indistinguishability of the bulk dynamics that forces resolvents to converge strongly in the far field-region.
We will later argue (in Sec.~\ref{sec:limsup}) that, in regions where the strong resolvent convergence
\[
(z-\hat{H}_{\Omega_N})^{-1}\rightarrow (z-\hat{H})^{-1}
\]
holds, one also expects the local GF, Eq.~(\ref{eq:local_GF_def}), to coincide in the thermodynamic limit with the trace of the resolvent, Eq.~(\ref{eq:resolvent-trace}).

Rephrased in terms of the trace of the resolvent, this means that the far-field values of $G_\Omega(z,\bar z)$ are determined by the bulk dynamics of the system. Since the bulk dynamics is expected to be independent of boundary conditions in the thermodynamic limit, $G_\Omega(z,\bar z)$ should coincide with $G(z)$ in the far-field region, i.e., for sufficiently large finite $|z|$.

\section{Hermitization and the Brown measure}

\subsection{Construction of the DOS from Hermitized Hamiltonian}\label{sec:brown_measure}
While, as already mentioned, there are in principle many ways to define the DOS of the infinite non-Hermitian system, it is convenient to have a practical instance of the class of equivalent ones, which is unambiguously defined for a broad class of systems and does not depend on the specific limiting procedure used to realize the infinite-system limit.

Motivated by the previous section, we can demand that any such definition $\rho_{\mathrm{B}}(z,\bar z)$ should produce the same far-field configuration of its corresponding Green's function $$ G_{\mathrm{B}}(z,\bar z)=\int_\mathbb{C}\mathrm{d}^2\omega\ \rho_\mathrm{B} (\omega,\bar \omega) (z-\omega)^{-1}$$ as the true resolvent $G(z)=\bra{i}(z-\hat{H})^{-1}\ket{i}$ of the Hamiltonian of the infinite system does.
It would also be convenient if such $G_{\mathrm{B}}(z,\bar z)$ coincides with $G(z)$ not just in the far-field region, but everywhere where
$(z-\hat{H})^{-1}$ is defined.

In other words, we expect $G_{\mathrm{B}}(z,\bar z)$ to continue $G(z)$ on the spectral set, where $(z-\hat{H})^{-1}$ is not defined. Alternatively, we expect $\rho_{\mathrm{B}}(z,\bar z)=\frac{1}{\pi}\bar \partial G_{\mathrm{B}}(z,\bar z)$ to be supported on the spectral set of the actually infinite operator $\hat H$: a convenient property, that would allow us to restore the spectrum from the corresponding DOS.

Such a ``spectral density'' can actually be straightforwardly constructed through the so-called ``Hermitization'' procedure, and this object was long ago introduced in various contexts, e.g., in the theory of non-Hermitian random matrices~\cite{Feinberg:1997}.
Formally, such an object is known in mathematical literature as the Brown measure of a (generally non-normal) bounded operator in a tracial von Neumann algebra~\cite{brown1983lidskii}.

We will shortly describe its construction as well as the practically relevant intuition behind its properties; the fully rigorous treatment can be found in the operator-algebra literature.

To proceed, we first define the ``Hermitized'' Hamiltonian:
	\begin{equation}\label{eq:Infinite_Hermitized_H}
		\hat{\mathbb{H}}(z)
		=
		\begin{pmatrix}
			0 & \hat H - z \\
			\hat H^\dagger - \bar z & 0
		\end{pmatrix}.
	\end{equation}
Notice that $z$ is kept here as a bookkeeping tool. 
For all $z$ away from the spectrum of $\hat H$, the inverse of the Hermitian operator $\hat{\mathbb{H}}(z)$ is well defined and is given by:  
\begin{equation} \label{eq:Infinite_Hermitized_H_inv}
	-\hat{\mathbb{H}}^{-1}(z)
	=
	\begin{pmatrix}
		0 &  ( \bar z-\hat H^\dagger )^{-1}\\
		 (z-\hat H )^{-1}& 0
	\end{pmatrix}.
\end{equation}
Thus, whenever \(z\) belongs to the resolvent set of \(\hat H\), the Hermitized Hamiltonian has a spectral gap at zero energy. 
Conversely, if \(z\) belongs to the spectrum of \(\hat H\), then \(\hat{\mathbb{H}}(z)\) is not invertible. 
We therefore obtain the useful equivalence
\begin{equation}
	z\in\operatorname{spec}(\hat H)\quad\Leftrightarrow\quad0\in\operatorname{spec}[\hat{\mathbb{H}}(z)].
\end{equation}
In what follows, we suppress the explicit dependence on \(z\) and write \(\hat{\mathbb{H}}\equiv\hat{\mathbb{H}}(z)\) whenever no ambiguity can arise.

Equation~(\ref{eq:Infinite_Hermitized_H}) implies that for all $z$ away from the spectrum of $\hat{H}$, projecting onto the lower left block of $\hat{\mathbb{H}}$ recovers the original resolvent, i.e., 
\begin{equation}
	-\bra{i,2}\hat{\mathbb{H}}^{-1}\ket{i,1}=\bra{i}(z-\hat H)^{-1}\ket{i}=G(z),
\end{equation} 
where the second index in the state notation denotes the Hermitian double subspace label. 
Equivalently, such inverse can be recovered through a boundary value of the resolvent of the \textit{Hermitian} operator $\hat{\mathbb{H}}$, $\hat{\mathbb{G}}(\omega)\equiv(\omega - \hat{\mathbb{H}})^{-1}$, so that: 
\begin{equation}\label{eq:GF_to_HD_GF}
    	(z-\hat H)^{-1}=\lim_{\eta\rightarrow0^+}\hat{\mathbb{G}}(0+i\eta)\big|_{2,1}
\end{equation}
where $\big|_{2,1}$ denotes the $(2,1)$ block of $\hat{\mathbb{G}}(0+i\eta)$ in the Hermitian double basis, i.e.,
$$
\hat{\mathbb{G}}(0+i\eta)\big|_{2,1}
=
P_2\,\hat{\mathbb{G}}(0+i\eta)\,P_1,
$$
with projectors
$
P_\alpha=\sum_i |i,\alpha\rangle\langle i,\alpha|,\ \alpha\in\{1,2\}.
$
The limiting form (\ref{eq:GF_to_HD_GF}) is guaranteed to work as long as $0$ is not in the spectrum of $\hat{\mathbb{H}}$, i.e., for all $z$ outside of the spectrum of $\hat H$. 

Outside of the spectrum of $\hat H$ one is thus able to compute its local resolvent $G(z)$ as a certain local GF's matrix element of the Hermitized problem, i.e.:
\begin{equation}\label{eqn:Hermitized-inter-site-GF}
    G(z)=\lim_{\eta\rightarrow0^+}\bra{i,2}\hat{\mathbb{G}}(0+i\eta)\ket{i,1},\quad z\notin \operatorname{spec}(\hat H).
\end{equation}
While this equality holds on the resolvent set of $\hat H$ (i.e., in the complement of the spectrum), the right-hand side is a well-defined quantity even for those $z$ inside the spectrum (as just a boundary value of some inter-site Green's function of the Hermitian problem). 
We can therefore treat the right-hand side of the last expression as the definition of the (one possible) extension of $\bra{i}(z-\hat H)^{-1}\ket{i}$ to the whole $\mathbb{C}$; thus, we define:
\begin{equation}\label{eq:Brown_GF}
    G_\mathrm{B}(z,\bar z)= \lim_{\eta\rightarrow0^+}\bra{i,2}\hat{\mathbb{G}}(0+i\eta)\ket{i,1},\quad \forall z\in \mathbb C
\end{equation}
with the property that $G_\mathrm{B}(z,\bar z)=G(z)$ on $z$ outside the spectrum of $\hat H$.

One now defines the Brown spectral measure as
\begin{equation}\label{eq:Brown_measure}
    \rho_\mathrm{B}(z,\bar z)=\frac{1}{\pi}\bar \partial G_\mathrm{B}(z,\bar z)
\end{equation}
which is supported only on the spectrum of $\hat H$ [as $G_\mathrm{B}(z,\bar z)$ defines an analytic function outside of it].\footnote{For the sake of our discussion, note that it is guaranteed that Eq.~(\ref{eq:Brown_measure}) defines the density of a positive Radon measure on $\mathbb C$ if $\hat H$ belongs to a tracial von Neumann algebra. This applies to arbitrary translationally invariant hopping Hamiltonians on flat lattices, but also holds in more general settings.}

By no means is the Brown measure the canonically unique spectral measure in the aforementioned physical sense, namely in the sense that all the physical information is encoded in the multipole moments, i.e., the far fields, rather than in the precise form of the limiting spectral distribution. The Brown measure is only one such choice of distribution possessing these far fields. It is, however, a convenient object, having the property of being supported only on the spectrum of the infinite operator, as well as coinciding with the usual spectral measure in the Hermitian case.

In the systems of current interest, namely, Euclidean lattice systems, the Brown measure is also expected to describe the asymptotic eigenvalue distribution of PBC clusters. This can be seen from the definition in Eq.~(\ref{eq:Brown_GF}). Upon passing to the Fourier basis of $\hat{\mathbb{H}}$, such that $\hat{H}$ is transformed into a Bloch Hamiltonian $h(k)$, one finds by direct calculation
\[
G_\mathrm{B}(z,\bar z) = -\int_{\mathrm{B.Z.}}\frac{\bar z - \bar h(k)}{(i0^+)^2-[z-h(k)][\bar z - \bar h(k)]},
\]
which simplifies to
\[
G_\mathrm{B}(z,\bar z) = \int_{\mathrm{B.Z.}}\frac{1}{z-h(k)}.
\]
This is precisely the limiting form of the trace of the resolvent for PBC clusters, which allows us to conclude that
\begin{equation}
    G_\mathrm{B}(z,\bar z)=G_\mathrm{PBC}(z,\bar z).
\end{equation}

This derivation extends straightforwardly to lattices with a nontrivial unit cell, in which case $h(k)$ becomes matrix-valued and the spectral density is recovered by taking the appropriately normalized trace over the unit-cell degrees of freedom.

It is also important to note that approximations to the Brown measure can, in principle, be obtained from finite flakes with arbitrary boundary conditions; see Eq.~(\ref{eq:Brown_GF_from_finite_flakes}) below.

\subsection{Which thermodynamic limits produce \texorpdfstring{\nopagebreak\\}{} spectral distributions different from $\rho_{\mathrm{B}}(z,\bar z)$?}\label{sec:limsup}
The discussion in the previous subsection, in particular the construction of the Brown measure, concerned only the infinite operator associated with the infinite lattice. In that construction, the result depended solely on the limiting operator itself and not on the specific sequence of finite truncations used to approximate it.

To address the question posed in the present subsection, however, one must return to the conventional definition of the spectral density in terms of the limiting trace of the resolvent, Eq.~(\ref{eq:resolvent-trace}). Our goal here is to relate the corresponding limiting Green's functions \(G_\Omega\) to the intrinsic Green's function \(G_{\mathrm B}\) introduced in the previous subsection. In particular, we will argue that in the far-field region the two coincide,
\[
G_\Omega(z,\bar z)=G_{\mathrm B}(z,\bar z),
\]
which, per the discussion in Sec.~\ref{sec:brown_measure}, further implies that $G_\Omega(z,\bar z)$ coincides with $G(z)=\bra{i}(z-\hat H)^{-1}\ket{i}$ for sufficiently large complex frequencies. In particular, $G_\Omega(z, \bar z)$ in that region is given by the multipole expansion in Eq.~(\ref{eq:multipole_expansion}).

To understand under which conditions a given thermodynamic limit produces a spectral measure distinct from the Brown measure, we once again employ the Hermitization procedure. For a given approximating sequence of Hamiltonians \(\hat H_{\Omega_N}\), with \(\hat H_{\Omega_N}\to \hat H\) strongly, we define the Hermitian doubled Hamiltonian
\begin{equation}
\label{eq:Hermitian_Doubled_Hamiltonian}
\hat{\mathbb{H}}_{\Omega_N}
=
\begin{pmatrix}
0 & \hat H_{\Omega_N}-z\\
\hat H_{\Omega_N}^\dagger-\bar z & 0
\end{pmatrix}.
\end{equation}
We further introduce its resolvent
\begin{equation}
\label{eq:resolvent_of_Hermitian_double}
\hat{\mathbb G}_{\Omega_N}(\omega)
=
(\omega-\hat{\mathbb{H}}_{\Omega_N})^{-1}.
\end{equation}
Since each \(\hat{\mathbb{H}}_{\Omega_N}\) is Hermitian, strong convergence \(\hat{\mathbb{H}}_{\Omega_N}\to \hat{\mathbb{H}}\) implies strong resolvent convergence at complex frequencies,
\[
\hat{\mathbb G}_{\Omega_N}(\omega)\to \hat{\mathbb G}(\omega),
\qquad
\omega\in\mathbb C\setminus\mathbb R.
\]

As a consequence, the Green's function \(G_{\mathrm B}(z,\bar z)\) of the infinite system can be obtained from any sequence of finite truncations as
\begin{equation}\label{eq:Brown_GF_from_finite_flakes}
G_{\mathrm B}(z,\bar z)
=
\lim_{\eta\to0^+}\lim_{N\to\infty}
\bra{i,2}\hat{\mathbb G}_{\Omega_N}(i\eta)\ket{i,1},
\end{equation}
that is, with the thermodynamic limit taken before the zero-frequency limit. For any fixed \(\eta>0\), the dependence on the choice of boundary conditions disappears in the thermodynamic limit as a consequence of SRC. Equation~\eqref{eq:Brown_GF_from_finite_flakes} therefore provides a practical way to approximate the Brown measure from finite flakes with arbitrary boundary conditions, provided the regulator \(\eta\) is kept finite until after the thermodynamic limit is taken.

At finite \(\eta\), the right-hand side of Eq.~\eqref{eq:Brown_GF_from_finite_flakes} is simply a local Green's function matrix element of a \emph{Hermitian} problem. 
For a translationally invariant lattice, such a local quantity is expected to coincide in the thermodynamic limit with the corresponding spatial average, since the boundary contribution becomes negligible. One therefore expects
\begin{equation}\label{eq:Brown_GF_from_finite_flakes_trace}
G_{\mathrm B}(z,\bar z)
=
\lim_{\eta\to0^+}\lim_{N\to\infty}
\frac{1}{N}\sum_i
\bra{i,2}\hat{\mathbb G}_{\Omega_N}(i\eta)\ket{i,1},
\end{equation}
where the sum runs over all lattice sites contained in the flake~\(\Omega_N\).

By contrast, when one studies the limiting spectral distribution associated with a \emph{specific} choice of boundary conditions, the relevant object is the trace of the resolvent, Eq.~(\ref{eq:resolvent-trace}). In terms of the Hermitian double it is given by
\begin{equation}\label{eq:arbitrary_BC_GF_from_finite_flakes}
G_\Omega(z,\bar z)
=
\lim_{N\to\infty}
\frac{1}{N}\sum_i
\bra{i,2}\hat{\mathbb G}_{\Omega_N}(0)\ket{i,1},
\end{equation}
that is, by the same spatial average, but with the zero-frequency limit taken before the thermodynamic limit. The possibility that different thermodynamic limits yield spectral measures different from the Brown measure is therefore a manifestation of the non-commutativity of the limits
\[
\eta\to0^+,
\qquad
N\to\infty.
\]
This is analogous in spirit to other familiar non-commuting limits in many-body physics, such as the hydrodynamic and collisionless limits in Fermi-liquid theory~\cite{pines2018theory}. 

The comparison between Eqs.~\eqref{eq:Brown_GF_from_finite_flakes_trace} and \eqref{eq:arbitrary_BC_GF_from_finite_flakes} makes it possible to identify when \(G_\Omega\) and \(G_{\mathrm B}\) coincide. Since \(\hat{\mathbb G}_{\Omega_N}(\omega)\) is the resolvent of a chiral-symmetric Hermitian Hamiltonian, the usual spectral decomposition applies. For a given \(z\), there are two qualitatively distinct possibilities: either \(0\in\operatorname{spec}(\hat{\mathbb{H}})\) or \(0\notin\operatorname{spec}(\hat{\mathbb{H}})\). Equivalently, either \(z\in\operatorname{spec}(\hat H)\) or \(z\) belongs to the resolvent set of \(\hat H\).

If \(0\in\operatorname{spec}(\hat{\mathbb{H}})\), then the poles of \(\hat{\mathbb G}_{\Omega_N}(\omega)\) are expected to accumulate near \(\omega=0\) as \(N\to\infty\). In that case, continuity in \(\eta\) at \(\eta=0\) is generally lost, and the two limits need not agree. Consequently, \(G_\Omega(z,\bar z)\) and \(G_{\mathrm B}(z,\bar z)\) may differ.

If, on the other hand, \(0\notin\operatorname{spec}(\hat{\mathbb{H}})\) [so that $z\notin \operatorname{spec}(\hat H)$], then the infinite Hermitized system is gapped at zero. One then expects the poles of \(\hat{\mathbb G}_{\Omega_N}(\omega)\) to remain away from a finite neighborhood of \(\omega=0\), so that the dependence on \(\eta\) remains regular near zero. In that case the two limits are expected to commute, leading to
\[
G_\Omega(z,\bar z)=G_{\mathrm B}(z,\bar z).
\]
An analogous argument, together with the reasoning used to pass from \eqref{eq:Brown_GF_from_finite_flakes} to \eqref{eq:Brown_GF_from_finite_flakes_trace}, allows one to conclude that, for such \( z \), the limiting local Green's function coincides with the trace of the resolvent. Namely, for such \( z \), one generally has 
\[
G_\Omega(z,\bar z)=G_\Omega(i,i;z).
\]

There is, however, an important exception. 
A gap in the spectrum of the infinite Hamiltonian \(\hat{\mathbb{H}}\) at zero does \emph{not} by itself guarantee that the finite truncations \(\hat{\mathbb{H}}_{\Omega_N}\) do not possess eigenvalues arbitrarily close to zero. This may happen when the chosen sequence of approximants is exceptional, in the sense that it supports near-zero modes that are absent in the thermodynamic limit. A simple example is provided by boundary states whose energies approach zero as \(N\to\infty\). Such a mechanism is robust when the Hermitized system realizes a nontrivial topological insulating phase, which enforces topological zero modes by means of the bulk-boundary correspondence.

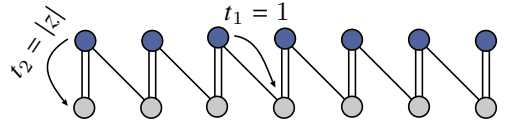
\begin{figure}
\hspace{-0.7 cm}
    \centering
\begin{tikzpicture}[x=0.75pt,y=0.75pt,yscale=-1.4,xscale=1.4, scale=0.8]
\draw  [line width=.6]  (242.13,14.13) -- (242.13,44.13)(239.13,14.13) -- (239.13,44.13) ;
\draw   [line width=.6] (272.13,14.13) -- (272.13,44.13)(269.13,14.13) -- (269.13,44.13) ;
\draw  [line width=.6]  (301.63,13.63) -- (301.63,43.63)(298.63,13.63) -- (298.63,43.63) ;
\draw [line width=.6]   (331.63,14.13) -- (331.63,44.13)(328.63,14.13) -- (328.63,44.13) ;
\draw  [line width=.6]  (361.63,13.63) -- (361.63,43.63)(358.63,13.63) -- (358.63,43.63) ;
\draw  [line width=.6]  (391.63,13.63) -- (391.63,43.63)(388.63,13.63) -- (388.63,43.63) ;
\draw  [line width=.6]  (421.63,14.13) -- (421.63,44.13)(418.63,14.13) -- (418.63,44.13) ;
\draw  [line width=.6]  (240.63,14.13) -- (270.63,44.13) ;
\draw  [line width=.6]  (270.63,14.13) -- (300.63,44.13) ;
\draw  [line width=.6]  (300.13,14.13) -- (330.13,44.13) ;
\draw  [line width=.6]  (330.13,13.63) -- (360.13,43.63) ;
\draw  [line width=.6]  (360.13,13.63) -- (390.13,43.63) ;
\draw  [line width=.6]  (390.13,13.63) -- (420.13,43.63) ;
\draw  [fill={rgb, 255:red, 202; green, 202; blue, 202 }  ,fill opacity=1 ] [line width=.6] (265.5,44.13) .. controls (265.5,41.57) and (267.57,39.5) .. (270.13,39.5) .. controls (272.68,39.5) and (274.75,41.57) .. (274.75,44.13) .. controls (274.75,46.68) and (272.68,48.75) .. (270.13,48.75) .. controls (267.57,48.75) and (265.5,46.68) .. (265.5,44.13) -- cycle ;
\draw  [fill={rgb, 255:red, 202; green, 202; blue, 202 }  ,fill opacity=1 ] [line width=.6] (235.5,44.13) .. controls (235.5,41.57) and (237.57,39.5) .. (240.13,39.5) .. controls (242.68,39.5) and (244.75,41.57) .. (244.75,44.13) .. controls (244.75,46.68) and (242.68,48.75) .. (240.13,48.75) .. controls (237.57,48.75) and (235.5,46.68) .. (235.5,44.13) -- cycle ;
\draw  [fill={rgb, 255:red, 202; green, 202; blue, 202 }  ,fill opacity=1 ][line width=.6] (295,43.63) .. controls (295,41.07) and (297.07,39) .. (299.63,39) .. controls (302.18,39) and (304.25,41.07) .. (304.25,43.63) .. controls (304.25,46.18) and (302.18,48.25) .. (299.63,48.25) .. controls (297.07,48.25) and (295,46.18) .. (295,43.63) -- cycle ;
\draw  [fill={rgb, 255:red, 202; green, 202; blue, 202 }  ,fill opacity=1 ][line width=.6] (355.5,43.63) .. controls (355.5,41.07) and (357.57,39) .. (360.13,39) .. controls (362.68,39) and (364.75,41.07) .. (364.75,43.63) .. controls (364.75,46.18) and (362.68,48.25) .. (360.13,48.25) .. controls (357.57,48.25) and (355.5,46.18) .. (355.5,43.63) -- cycle ;
\draw  [fill={rgb, 255:red, 202; green, 202; blue, 202 }  ,fill opacity=1 ][line width=.6] (325,44.13) .. controls (325,41.57) and (327.07,39.5) .. (329.63,39.5) .. controls (332.18,39.5) and (334.25,41.57) .. (334.25,44.13) .. controls (334.25,46.68) and (332.18,48.75) .. (329.63,48.75) .. controls (327.07,48.75) and (325,46.68) .. (325,44.13) -- cycle ;
\draw  [fill={rgb, 255:red, 202; green, 202; blue, 202 }  ,fill opacity=1 ][line width=.6] (415.5,44.13) .. controls (415.5,41.57) and (417.57,39.5) .. (420.13,39.5) .. controls (422.68,39.5) and (424.75,41.57) .. (424.75,44.13) .. controls (424.75,46.68) and (422.68,48.75) .. (420.13,48.75) .. controls (417.57,48.75) and (415.5,46.68) .. (415.5,44.13) -- cycle ;
\draw  [fill={rgb, 255:red, 202; green, 202; blue, 202 }  ,fill opacity=1 ][line width=.6] (385.5,43.63) .. controls (385.5,41.07) and (387.57,39) .. (390.13,39) .. controls (392.68,39) and (394.75,41.07) .. (394.75,43.63) .. controls (394.75,46.18) and (392.68,48.25) .. (390.13,48.25) .. controls (387.57,48.25) and (385.5,46.18) .. (385.5,43.63) -- cycle ;
\draw  [fill={rgb, 255:red, 81; green, 100; blue, 155 }  ,fill opacity=1 ][line width=.6] (266,14.13) .. controls (266,11.57) and (268.07,9.5) .. (270.63,9.5) .. controls (273.18,9.5) and (275.25,11.57) .. (275.25,14.13) .. controls (275.25,16.68) and (273.18,18.75) .. (270.63,18.75) .. controls (268.07,18.75) and (266,16.68) .. (266,14.13) -- cycle ;
\draw  [fill={rgb, 255:red, 81; green, 100; blue, 155 }  ,fill opacity=1 ][line width=.6] (236,14.13) .. controls (236,11.57) and (238.07,9.5) .. (240.63,9.5) .. controls (243.18,9.5) and (245.25,11.57) .. (245.25,14.13) .. controls (245.25,16.68) and (243.18,18.75) .. (240.63,18.75) .. controls (238.07,18.75) and (236,16.68) .. (236,14.13) -- cycle ;
\draw  [fill={rgb, 255:red, 81; green, 100; blue, 155 }  ,fill opacity=1 ][line width=.6] (295.5,12.63) .. controls (295.5,10.07) and (297.57,8) .. (300.13,8) .. controls (302.68,8) and (304.75,10.07) .. (304.75,12.63) .. controls (304.75,15.18) and (302.68,17.25) .. (300.13,17.25) .. controls (297.57,17.25) and (295.5,15.18) .. (295.5,12.63) -- cycle ;
\draw  [fill={rgb, 255:red, 81; green, 100; blue, 155 }  ,fill opacity=1 ][line width=.6] (355.5,13.63) .. controls (355.5,11.07) and (357.57,9) .. (360.13,9) .. controls (362.68,9) and (364.75,11.07) .. (364.75,13.63) .. controls (364.75,16.18) and (362.68,18.25) .. (360.13,18.25) .. controls (357.57,18.25) and (355.5,16.18) .. (355.5,13.63) -- cycle ;
\draw  [fill={rgb, 255:red, 81; green, 100; blue, 155 }  ,fill opacity=1 ][line width=.6] (325.5,13.13) .. controls (325.5,10.57) and (327.57,8.5) .. (330.13,8.5) .. controls (332.68,8.5) and (334.75,10.57) .. (334.75,13.13) .. controls (334.75,15.68) and (332.68,17.75) .. (330.13,17.75) .. controls (327.57,17.75) and (325.5,15.68) .. (325.5,13.13) -- cycle ;
\draw  [fill={rgb, 255:red, 81; green, 100; blue, 155 }  ,fill opacity=1 ][line width=.6] (415.5,14.13) .. controls (415.5,11.57) and (417.57,9.5) .. (420.13,9.5) .. controls (422.68,9.5) and (424.75,11.57) .. (424.75,14.13) .. controls (424.75,16.68) and (422.68,18.75) .. (420.13,18.75) .. controls (417.57,18.75) and (415.5,16.68) .. (415.5,14.13) -- cycle ;
\draw  [fill={rgb, 255:red, 81; green, 100; blue, 155 }  ,fill opacity=1, line width=0.5 ][line width=.6] (385.5,13.63) .. controls (385.5,11.07) and (387.57,9) .. (390.13,9) .. controls (392.68,9) and (394.75,11.07) .. (394.75,13.63) .. controls (394.75,16.18) and (392.68,18.25) .. (390.13,18.25) .. controls (387.57,18.25) and (385.5,16.18) .. (385.5,13.63) -- cycle ;
\draw  [line width=.6]  (232.8,13.32) .. controls (218.8,22.04) and (223.01,30.96) .. (230.95,41.03) ;
\draw [shift={(232.8,43.32)}, rotate = 230.61] [fill={rgb, 255:red, 0; green, 0; blue, 0 }  ][line width=0.6]  [draw opacity=0] (2.93,-1.29) -- (0,0) -- (2.93,1.29) -- cycle    ;
\draw  [line width=.6]  (307.35,11.9) .. controls (312.48,12.56) and (315.8,15.36) .. (318.3,19.06) .. controls (320.93,22.97) and (322.63,27.89) .. (324.54,32.4) ;
\draw [shift={(325.75,35.1)}, rotate = 244.44] [fill={rgb, 255:red, 0; green, 0; blue, 0 }  ][line width=0.6]  [draw opacity=0] (2.93,-1.29) -- (0,0) -- (2.93,1.29) -- cycle    ;

\draw (202.73,26.8) node [anchor=north west][inner sep=0.75pt]  [font=\normalsize,rotate=-303.86]  {$t_{2} =|z|$};
\draw (301.8,-4) node [anchor=north west][inner sep=0.75pt]  [font=\normalsize]  {$t_{1} =1$};

\end{tikzpicture}
    \caption{Pictorial representation of the Hermitian double of the unidirectional Hatano-Nelson model, Eq.~(\ref{eq:Hermitized_unidir_HN}). Blue and grey sites denote the sublattice degree of freedom introduced by the Hermitian doubling. Single-line bonds represent (reciprocal) hopping amplitudes $t_1 =1$, while double-line bonds correspond to $t_2 = |z|$. }
    \label{fig:HN_Hermitian_Double}
\end{figure}

We are thus led to the following general expectation: 
\begin{equation}\label{eq:universal_limit}
G_\Omega(z,\bar z)=G_\Omega(i,i;z)=G_{\mathrm B}(z,\bar z)
\end{equation}
for all \(z\) such that the corresponding Hermitized system is neither gapless at zero nor topologically nontrivial in a way that produces boundary-localized near-zero modes in the chosen truncation sequence.\footnote{
Note that, for a given \( z \), Eq.~(\ref{eq:universal_limit}) is guaranteed to hold whenever the strong resolvent convergence
\[
(z-\hat{H}_{\Omega_N})^{-1}\to (z-\hat{H})^{-1}
\]
is ensured. Indeed, in this case \(\hat{\mathbb{H}}_{\Omega_N}^{-1}\) also converges strongly to \(\hat{\mathbb{H}}^{-1}\), and hence
\[
\sup_N \norm{\hat{\mathbb{H}}_{\Omega_N}^{-1}} < \infty
\]
by the uniform boundedness principle. The latter condition implies the absence of zero-energy modes, so the assumptions required for Eq.~(\ref{eq:universal_limit}) are satisfied.
}

We can now justify the statement used repeatedly throughout the main text, namely that for sufficiently large \(|z|\) one always has
\[
G_\Omega(z,\bar z)=G(z).
\]
Indeed, for \(|z|>\|\hat H\|\), the Hermitized Hamiltonian~\eqref{eq:Infinite_Hermitized_H} is adiabatically connected to the trivial limit
\[
\hat{\mathbb{H}}_0
=
-
\begin{pmatrix}
0 & z\\
\bar z & 0
\end{pmatrix},
\]
and therefore does not support topological zero-energy boundary modes. In particular, the Hermitized problem is then both gapped at zero and topologically trivial, so the two limits commute due to \eqref{eq:universal_limit} and \(G_\Omega(z,\bar z)=G(z)\).

The preceding argument implicitly depends on the bulk-boundary correspondence, in that we treat trivial topology of the bulk gap as being equivalent to the absence of zero-energy modes (which is not necessarily the case, see Sec.~\ref{sec:energy_accumulation}). 
Strictly speaking, the precise bound depends on the approximating sequence \(\hat H_N\). If \(|z|>\|\hat H_N\|\), then the corresponding Hermitized Hamiltonian \(\hat{\mathbb{H}}_N\) has a gap at zero satisfying
\[
\Delta_N\ge 2\bigl(|z|-\|\hat H_N\|\bigr).
\]
Accordingly, the above conclusion is strictly justified whenever
\[
|z|>\limsup_{N\to\infty}\|\hat H_N\|.
\]
Notice that this coincides with the region where the Neumann series, Eq.~(\ref{eq:neumann_series}), applies for the given approximating sequence $\Omega_N$, and where strong resolvent convergence is guaranteed, see Sec.~\ref{sec:NH_constraints}.

\begin{figure}
    \centering

\begin{tikzpicture}[x=0.75pt,y=0.75pt,yscale=-1,xscale=1]
\draw  [color={rgb, 255:red, 0; green, 0; blue, 0 }  ,draw opacity=1 ][fill={rgb, 255:red, 255; green, 255; blue, 255 }  ,fill opacity=1 ] (197.17,5.86) -- (355.39,5.86) -- (355.39,165.16) -- (197.17,165.16) -- cycle ;
\draw  [color={rgb, 255:red, 58; green, 178; blue, 224 }  ,draw opacity=0 ][fill={rgb, 255:red, 235; green, 235; blue, 235 }  ,fill opacity=1 ][line width=2.25]  (217.22,85.51) .. controls (217.22,52.89) and (243.66,26.44) .. (276.28,26.44) .. controls (308.9,26.44) and (335.35,52.89) .. (335.35,85.51) .. controls (335.35,118.13) and (308.9,144.57) .. (276.28,144.57) .. controls (243.66,144.57) and (217.22,118.13) .. (217.22,85.51) -- cycle ;
\draw    (276.37,165.07) -- (276.37,7.66) ;
\draw [shift={(276.37,5.66)}, rotate = 90.28] [fill={rgb, 255:red, 0; green, 0; blue, 0 }  ][line width=0.08]  [draw opacity=0] (9.6,-2.4) -- (0,0) -- (9.6,2.4) -- cycle    ;
\draw    (197.38,85.06) -- (353.5,85.06) ;
\draw [shift={(355.5,85.06)}, rotate = 180.2] [fill={rgb, 255:red, 0; green, 0; blue, 0 }  ][line width=0.08]  [draw opacity=0] (9.6,-2.4) -- (0,0) -- (9.6,2.4) -- cycle    ;
\draw  [fill={rgb, 255:red, 255; green, 255; blue, 255 }  ,fill opacity=1 ] (396.97,36.69) -- (493.03,36.69) -- (493.03,76.69) -- (396.97,76.69) -- cycle ;
\draw  [fill={rgb, 255:red, 255; green, 255; blue, 255 }  ,fill opacity=1 ] (396.97,98.19) -- (493.03,98.19) -- (493.03,138.19) -- (396.97,138.19) -- cycle ;
\draw    (346.05,118.49) -- (394.05,118.49) ;
\draw [shift={(346.05,118.49)}, rotate = 0] [color={rgb, 255:red, 0; green, 0; blue, 0 }  ][fill={rgb, 255:red, 0; green, 0; blue, 0 }  ][line width=0.75]      (0, 0) circle [x radius= 1.34, y radius= 1.34]   ;
\draw  [color={rgb, 255:red, 58; green, 80; blue, 126 }  ,draw opacity=1 ][fill={rgb, 255:red, 255; green, 255; blue, 255 }  ,fill opacity=0 ][line width=2.25]  (217.37,84.79) .. controls (217.37,52.17) and (243.82,25.72) .. (276.44,25.72) .. controls (309.06,25.72) and (335.5,52.17) .. (335.5,84.79) .. controls (335.5,117.41) and (309.06,143.85) .. (276.44,143.85) .. controls (243.82,143.85) and (217.37,117.41) .. (217.37,84.79) -- cycle ;
\draw    (306,56.7) -- (394,56.7) ;
\draw [shift={(306,56.7)}, rotate = 0] [color={rgb, 255:red, 0; green, 0; blue, 0 }  ][fill={rgb, 255:red, 0; green, 0; blue, 0 }  ][line width=0.75]      (0, 0) circle [x radius= 1.34, y radius= 1.34]   ;
\draw    (325.5,23.16) .. controls (318.63,24.52) and (330,27.16) .. (325,32.16) .. controls (320.15,37.01) and (321.89,20.22) .. (312.87,36.09) ;
\draw [shift={(312,37.66)}, rotate = 298.39] [color={rgb, 255:red, 0; green, 0; blue, 0 }  ][line width=0.5]    (6.56,-1.97) .. controls (4.17,-0.84) and (1.99,-0.18) .. (0,0) .. controls (1.99,0.18) and (4.17,0.84) .. (6.56,1.97)   ;
\draw    (398.5,56.7) -- (489.5,56.68) ;
\draw [shift={(491.5,56.68)}, rotate = 179.99] [fill={rgb, 255:red, 0; green, 0; blue, 0 }  ][line width=0.08]  [draw opacity=0] (7.2,-1.8) -- (0,0) -- (7.2,1.8) -- cycle    ;
\draw [shift={(445,56.69)}, rotate = 179.99] [color={rgb, 255:red, 0; green, 0; blue, 0 }  ][line width=0.75]    (0,3.35) -- (0,-3.35)   ;
\draw  [color={rgb, 255:red, 58; green, 80; blue, 126 }  ,draw opacity=0 ][fill={rgb, 255:red, 58; green, 80; blue, 126 }  ,fill opacity=1 ] (429.5,56.69) .. controls (429.5,55.86) and (428.83,55.19) .. (428,55.19) .. controls (427.17,55.19) and (426.5,55.86) .. (426.5,56.69) .. controls (426.5,57.52) and (427.17,58.19) .. (428,58.19) .. controls (428.83,58.19) and (429.5,57.52) .. (429.5,56.69) -- cycle ;
\draw  [color={rgb, 255:red, 58; green, 80; blue, 126 }  ,draw opacity=0 ][fill={rgb, 255:red, 58; green, 80; blue, 126 }  ,fill opacity=1 ] (426.5,56.69) .. controls (426.5,55.86) and (425.83,55.19) .. (425,55.19) .. controls (424.17,55.19) and (423.5,55.86) .. (423.5,56.69) .. controls (423.5,57.52) and (424.17,58.19) .. (425,58.19) .. controls (425.83,58.19) and (426.5,57.52) .. (426.5,56.69) -- cycle ;
\draw  [color={rgb, 255:red, 58; green, 80; blue, 126 }  ,draw opacity=0 ][fill={rgb, 255:red, 58; green, 80; blue, 126 }  ,fill opacity=1 ] (423.5,56.69) .. controls (423.5,55.86) and (422.83,55.19) .. (422,55.19) .. controls (421.17,55.19) and (420.5,55.86) .. (420.5,56.69) .. controls (420.5,57.52) and (421.17,58.19) .. (422,58.19) .. controls (422.83,58.19) and (423.5,57.52) .. (423.5,56.69) -- cycle ;
\draw  [color={rgb, 255:red, 58; green, 80; blue, 126 }  ,draw opacity=0 ][fill={rgb, 255:red, 58; green, 80; blue, 126 }  ,fill opacity=1 ] (420.17,56.69) .. controls (420.17,55.86) and (419.5,55.19) .. (418.67,55.19) .. controls (417.84,55.19) and (417.17,55.86) .. (417.17,56.69) .. controls (417.17,57.52) and (417.84,58.19) .. (418.67,58.19) .. controls (419.5,58.19) and (420.17,57.52) .. (420.17,56.69) -- cycle ;
\draw  [color={rgb, 255:red, 58; green, 80; blue, 126 }  ,draw opacity=0 ][fill={rgb, 255:red, 58; green, 80; blue, 126 }  ,fill opacity=1 ] (417.17,56.69) .. controls (417.17,55.86) and (416.5,55.19) .. (415.67,55.19) .. controls (414.84,55.19) and (414.17,55.86) .. (414.17,56.69) .. controls (414.17,57.52) and (414.84,58.19) .. (415.67,58.19) .. controls (416.5,58.19) and (417.17,57.52) .. (417.17,56.69) -- cycle ;
\draw  [color={rgb, 255:red, 58; green, 80; blue, 126 }  ,draw opacity=0 ][fill={rgb, 255:red, 58; green, 80; blue, 126 }  ,fill opacity=1 ] (414.17,56.69) .. controls (414.17,55.86) and (413.5,55.19) .. (412.67,55.19) .. controls (411.84,55.19) and (411.17,55.86) .. (411.17,56.69) .. controls (411.17,57.52) and (411.84,58.19) .. (412.67,58.19) .. controls (413.5,58.19) and (414.17,57.52) .. (414.17,56.69) -- cycle ;
\draw  [color={rgb, 255:red, 58; green, 80; blue, 126 }  ,draw opacity=0 ][fill={rgb, 255:red, 58; green, 80; blue, 126 }  ,fill opacity=1 ] (479.17,56.69) .. controls (479.17,55.86) and (478.5,55.19) .. (477.67,55.19) .. controls (476.84,55.19) and (476.17,55.86) .. (476.17,56.69) .. controls (476.17,57.52) and (476.84,58.19) .. (477.67,58.19) .. controls (478.5,58.19) and (479.17,57.52) .. (479.17,56.69) -- cycle ;
\draw  [color={rgb, 255:red, 58; green, 80; blue, 126 }  ,draw opacity=0 ][fill={rgb, 255:red, 58; green, 80; blue, 126 }  ,fill opacity=1 ] (476.17,56.69) .. controls (476.17,55.86) and (475.5,55.19) .. (474.67,55.19) .. controls (473.84,55.19) and (473.17,55.86) .. (473.17,56.69) .. controls (473.17,57.52) and (473.84,58.19) .. (474.67,58.19) .. controls (475.5,58.19) and (476.17,57.52) .. (476.17,56.69) -- cycle ;
\draw  [color={rgb, 255:red, 58; green, 80; blue, 126 }  ,draw opacity=0 ][fill={rgb, 255:red, 58; green, 80; blue, 126 }  ,fill opacity=1 ] (473.17,56.69) .. controls (473.17,55.86) and (472.5,55.19) .. (471.67,55.19) .. controls (470.84,55.19) and (470.17,55.86) .. (470.17,56.69) .. controls (470.17,57.52) and (470.84,58.19) .. (471.67,58.19) .. controls (472.5,58.19) and (473.17,57.52) .. (473.17,56.69) -- cycle ;
\draw  [color={rgb, 255:red, 58; green, 80; blue, 126 }  ,draw opacity=0 ][fill={rgb, 255:red, 58; green, 80; blue, 126 }  ,fill opacity=1 ] (469.83,56.69) .. controls (469.83,55.86) and (469.16,55.19) .. (468.33,55.19) .. controls (467.5,55.19) and (466.83,55.86) .. (466.83,56.69) .. controls (466.83,57.52) and (467.5,58.19) .. (468.33,58.19) .. controls (469.16,58.19) and (469.83,57.52) .. (469.83,56.69) -- cycle ;
\draw  [color={rgb, 255:red, 58; green, 80; blue, 126 }  ,draw opacity=0 ][fill={rgb, 255:red, 58; green, 80; blue, 126 }  ,fill opacity=1 ] (466.83,56.69) .. controls (466.83,55.86) and (466.16,55.19) .. (465.33,55.19) .. controls (464.5,55.19) and (463.83,55.86) .. (463.83,56.69) .. controls (463.83,57.52) and (464.5,58.19) .. (465.33,58.19) .. controls (466.16,58.19) and (466.83,57.52) .. (466.83,56.69) -- cycle ;
\draw  [color={rgb, 255:red, 58; green, 80; blue, 126 }  ,draw opacity=0 ][fill={rgb, 255:red, 58; green, 80; blue, 126 }  ,fill opacity=1 ] (463.83,56.69) .. controls (463.83,55.86) and (463.16,55.19) .. (462.33,55.19) .. controls (461.5,55.19) and (460.83,55.86) .. (460.83,56.69) .. controls (460.83,57.52) and (461.5,58.19) .. (462.33,58.19) .. controls (463.16,58.19) and (463.83,57.52) .. (463.83,56.69) -- cycle ;
\draw  [color={rgb, 255:red, 58; green, 80; blue, 126 }  ,draw opacity=0 ][fill={rgb, 255:red, 58; green, 80; blue, 126 }  ,fill opacity=1 ] (444,56.69) .. controls (444,55.86) and (443.33,55.19) .. (442.5,55.19) .. controls (441.67,55.19) and (441,55.86) .. (441,56.69) .. controls (441,57.52) and (441.67,58.19) .. (442.5,58.19) .. controls (443.33,58.19) and (444,57.52) .. (444,56.69) -- cycle ;
\draw  [color={rgb, 255:red, 58; green, 80; blue, 126 }  ,draw opacity=0 ][fill={rgb, 255:red, 58; green, 80; blue, 126 }  ,fill opacity=1 ] (449,56.69) .. controls (449,55.86) and (448.33,55.19) .. (447.5,55.19) .. controls (446.67,55.19) and (446,55.86) .. (446,56.69) .. controls (446,57.52) and (446.67,58.19) .. (447.5,58.19) .. controls (448.33,58.19) and (449,57.52) .. (449,56.69) -- cycle ;
\draw    (398.5,118.2) -- (489.5,118.18) ;
\draw [shift={(491.5,118.18)}, rotate = 179.99] [fill={rgb, 255:red, 0; green, 0; blue, 0 }  ][line width=0.08]  [draw opacity=0] (7.2,-1.8) -- (0,0) -- (7.2,1.8) -- cycle    ;
\draw [shift={(445,118.19)}, rotate = 179.99] [color={rgb, 255:red, 0; green, 0; blue, 0 }  ][line width=0.75]    (0,3.35) -- (0,-3.35)   ;
\draw  [color={rgb, 255:red, 58; green, 80; blue, 126 }  ,draw opacity=0 ][fill={rgb, 255:red, 58; green, 80; blue, 126 }  ,fill opacity=1 ] (429.5,118.19) .. controls (429.5,117.36) and (428.83,116.69) .. (428,116.69) .. controls (427.17,116.69) and (426.5,117.36) .. (426.5,118.19) .. controls (426.5,119.02) and (427.17,119.69) .. (428,119.69) .. controls (428.83,119.69) and (429.5,119.02) .. (429.5,118.19) -- cycle ;
\draw  [color={rgb, 255:red, 58; green, 80; blue, 126 }  ,draw opacity=0 ][fill={rgb, 255:red, 58; green, 80; blue, 126 }  ,fill opacity=1 ] (426.5,118.19) .. controls (426.5,117.36) and (425.83,116.69) .. (425,116.69) .. controls (424.17,116.69) and (423.5,117.36) .. (423.5,118.19) .. controls (423.5,119.02) and (424.17,119.69) .. (425,119.69) .. controls (425.83,119.69) and (426.5,119.02) .. (426.5,118.19) -- cycle ;
\draw  [color={rgb, 255:red, 58; green, 80; blue, 126 }  ,draw opacity=0 ][fill={rgb, 255:red, 58; green, 80; blue, 126 }  ,fill opacity=1 ] (423.5,118.19) .. controls (423.5,117.36) and (422.83,116.69) .. (422,116.69) .. controls (421.17,116.69) and (420.5,117.36) .. (420.5,118.19) .. controls (420.5,119.02) and (421.17,119.69) .. (422,119.69) .. controls (422.83,119.69) and (423.5,119.02) .. (423.5,118.19) -- cycle ;
\draw  [color={rgb, 255:red, 58; green, 80; blue, 126 }  ,draw opacity=0 ][fill={rgb, 255:red, 58; green, 80; blue, 126 }  ,fill opacity=1 ] (420.17,118.19) .. controls (420.17,117.36) and (419.5,116.69) .. (418.67,116.69) .. controls (417.84,116.69) and (417.17,117.36) .. (417.17,118.19) .. controls (417.17,119.02) and (417.84,119.69) .. (418.67,119.69) .. controls (419.5,119.69) and (420.17,119.02) .. (420.17,118.19) -- cycle ;
\draw  [color={rgb, 255:red, 58; green, 80; blue, 126 }  ,draw opacity=0 ][fill={rgb, 255:red, 58; green, 80; blue, 126 }  ,fill opacity=1 ] (417.17,118.19) .. controls (417.17,117.36) and (416.5,116.69) .. (415.67,116.69) .. controls (414.84,116.69) and (414.17,117.36) .. (414.17,118.19) .. controls (414.17,119.02) and (414.84,119.69) .. (415.67,119.69) .. controls (416.5,119.69) and (417.17,119.02) .. (417.17,118.19) -- cycle ;
\draw  [color={rgb, 255:red, 58; green, 80; blue, 126 }  ,draw opacity=0 ][fill={rgb, 255:red, 58; green, 80; blue, 126 }  ,fill opacity=1 ] (414.17,118.19) .. controls (414.17,117.36) and (413.5,116.69) .. (412.67,116.69) .. controls (411.84,116.69) and (411.17,117.36) .. (411.17,118.19) .. controls (411.17,119.02) and (411.84,119.69) .. (412.67,119.69) .. controls (413.5,119.69) and (414.17,119.02) .. (414.17,118.19) -- cycle ;
\draw  [color={rgb, 255:red, 58; green, 80; blue, 126 }  ,draw opacity=0 ][fill={rgb, 255:red, 58; green, 80; blue, 126 }  ,fill opacity=1 ] (479.17,118.19) .. controls (479.17,117.36) and (478.5,116.69) .. (477.67,116.69) .. controls (476.84,116.69) and (476.17,117.36) .. (476.17,118.19) .. controls (476.17,119.02) and (476.84,119.69) .. (477.67,119.69) .. controls (478.5,119.69) and (479.17,119.02) .. (479.17,118.19) -- cycle ;
\draw  [color={rgb, 255:red, 58; green, 80; blue, 126 }  ,draw opacity=0 ][fill={rgb, 255:red, 58; green, 80; blue, 126 }  ,fill opacity=1 ] (476.17,118.19) .. controls (476.17,117.36) and (475.5,116.69) .. (474.67,116.69) .. controls (473.84,116.69) and (473.17,117.36) .. (473.17,118.19) .. controls (473.17,119.02) and (473.84,119.69) .. (474.67,119.69) .. controls (475.5,119.69) and (476.17,119.02) .. (476.17,118.19) -- cycle ;
\draw  [color={rgb, 255:red, 58; green, 80; blue, 126 }  ,draw opacity=0 ][fill={rgb, 255:red, 58; green, 80; blue, 126 }  ,fill opacity=1 ] (473.17,118.19) .. controls (473.17,117.36) and (472.5,116.69) .. (471.67,116.69) .. controls (470.84,116.69) and (470.17,117.36) .. (470.17,118.19) .. controls (470.17,119.02) and (470.84,119.69) .. (471.67,119.69) .. controls (472.5,119.69) and (473.17,119.02) .. (473.17,118.19) -- cycle ;
\draw  [color={rgb, 255:red, 58; green, 80; blue, 126 }  ,draw opacity=0 ][fill={rgb, 255:red, 58; green, 80; blue, 126 }  ,fill opacity=1 ] (469.83,118.19) .. controls (469.83,117.36) and (469.16,116.69) .. (468.33,116.69) .. controls (467.5,116.69) and (466.83,117.36) .. (466.83,118.19) .. controls (466.83,119.02) and (467.5,119.69) .. (468.33,119.69) .. controls (469.16,119.69) and (469.83,119.02) .. (469.83,118.19) -- cycle ;
\draw  [color={rgb, 255:red, 58; green, 80; blue, 126 }  ,draw opacity=0 ][fill={rgb, 255:red, 58; green, 80; blue, 126 }  ,fill opacity=1 ] (466.83,118.19) .. controls (466.83,117.36) and (466.16,116.69) .. (465.33,116.69) .. controls (464.5,116.69) and (463.83,117.36) .. (463.83,118.19) .. controls (463.83,119.02) and (464.5,119.69) .. (465.33,119.69) .. controls (466.16,119.69) and (466.83,119.02) .. (466.83,118.19) -- cycle ;
\draw  [color={rgb, 255:red, 58; green, 80; blue, 126 }  ,draw opacity=0 ][fill={rgb, 255:red, 58; green, 80; blue, 126 }  ,fill opacity=1 ] (463.83,118.19) .. controls (463.83,117.36) and (463.16,116.69) .. (462.33,116.69) .. controls (461.5,116.69) and (460.83,117.36) .. (460.83,118.19) .. controls (460.83,119.02) and (461.5,119.69) .. (462.33,119.69) .. controls (463.16,119.69) and (463.83,119.02) .. (463.83,118.19) -- cycle ;

\draw (280.27,10.27) node [anchor=north west][inner sep=0.75pt]  [font=\footnotesize, scale=0.8]  {$\Im z$};
\draw (336.28,88.12) node [anchor=north west][inner sep=0.75pt]  [font=\footnotesize, scale=0.8]  {$\Re z$};
\draw (204.33,12.48) node [anchor=north west][inner sep=0.75pt]  [font=\footnotesize,scale=0.8]  {$G_{\mathrm{B}} = G_{\mathrm{OBC}}$};
\draw (454.17,124.4) node [anchor=north west][inner sep=0.75pt]  [font=\tiny]  {$\text{eigenvalues}$};
\draw (397,99.59) node [anchor=north west][inner sep=0.75pt]  [font=\tiny]  {$ \begin{array}{l}
\hat{\mathbb{H}}_{\mathrm{OBC_N}}\\
\\
|z| >1
\end{array}$};
\draw (277.57,92.83) node [anchor=north west, inner sep=0.75pt, font=\footnotesize, scale=0.8] {$G_{\mathrm{B}} \neq G_{\mathrm{OBC} }$};

\draw (397,37.59) node [anchor=north west][inner sep=0.75pt]  [font=\tiny]  {$ \begin{array}{l}
\hat{\mathbb{H}}_{\mathrm{OBC_{N}}}\\
\\
|z|< 1
\end{array}$};
\draw (454.11,62.2) node [anchor=north west][inner sep=0.75pt]  [font=\tiny]  {$\text{eigenvalues}$};
\draw (312,6.9) node [anchor=north west][inner sep=0.75pt]  [font=\footnotesize]  {$\operatorname{spec}(\hat{H})$};
\end{tikzpicture}

\caption{Two regimes of the Hermitized Hatano-Nelson model discussed in the main text: for $|z|>1$, the Hermitized OBC system is a trivial insulator. In this regime, the absence of zero modes guarantees that $G_{\mathrm{OBC}_N}(z)\to G_{\mathrm{B}}(z,\bar{z})=1/z$. For $|z|<1$, however, the presence of zero modes prevents one from using strong resolvent convergence to conclude that $G_{\mathrm{OBC}_N}(z)\to G_{\mathrm{B}}(z,\bar{z})$; in fact, in this regime $G_{\mathrm{B}}(z,\bar z)=0$, while $G_{\mathrm{OBC}_N}(z)\to 1/z$.}
    \label{fig:src_convergence_and_corner_modes}
\end{figure}
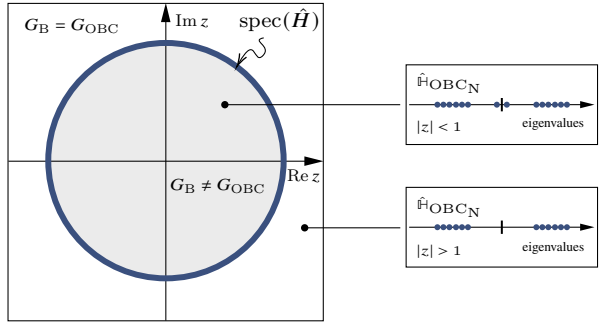

Previous considerations can be straightforwardly demonstrated, for example, in the case of the Hatano-Nelson model, Eq.~(\ref{eq:HN_Hamiltonian}). 
For simplicity, consider the unidirectional limit. The Hermitized Hamiltonian of the OBC Hatano-Nelson chain of size $N$ reads:
\begin{equation}\label{eq:Hermitized_unidir_HN}
    \hat{\mathbb{H}}_{\mathrm{OBC}_N}=\sum_{i=1}^{N-1}
        \ket{i+1,1}\bra{i,2}-\sum_{i=1}^Nz\ket{i,1}\bra{i,2}
    +
    h.c. 
\end{equation}
After a gauge transformation, \(z\) may be taken to be real and positive, \(z\to |z|\), so that \eqref{eq:Hermitized_unidir_HN} is equivalent to the 
Su-Schrieffer-Heeger (SSH) model~\cite{Su:1980}
with \(t_1=t_R=1\) and \(t_2=|z|\) (see Fig.~\ref{fig:HN_Hermitian_Double}). 
As is seen from Fig.~\ref{fig:HN_Hermitian_Double},  OBC truncation effectively cuts the $t_1$ bond of the Hermitized system. This bond is weak if $|z|>1$. There, the truncation is not topological and doesn't produce zero-energy edge modes; i.e., for $|z|>1$, $G_{\mathrm{OBC}}(z,\bar z)$ and $G_\mathrm{B}(z,\bar z)$ must agree: 
\begin{equation}
    G_{\mathrm{OBC}}(z,\bar z)=G_\mathrm{B}(z,\bar z),\quad |z|>1.
\end{equation}
For $|z|<1$, however, the truncation is topological; the presence of zero modes therefore prevents us from asserting that the limit~(\ref{eq:universal_limit}) holds.
As we know, there, in fact, the OBC and the PBC (Brown) Green's functions do not agree. This situation is illustrated schematically in Fig.~\ref{fig:src_convergence_and_corner_modes}.

Clearly, similar picture persists in more general settings; one can see that those perturbations which \textit{destroy} the topological modes of the Hermitized Hamiltonian will effectively prevent any limiting distributions---other than the Brown measure---from emerging in the thermodynamic limit. 
The simplest example here is adding a 
semi-transparent hopping between the last and the first sites in the HN chain in Eq.~(\ref{eq:HN_Hamiltonian}). Such hopping effectively interpolates between PBC and OBC. This perturbation will hybridize edge modes of the Hermitized Hamiltonian, gapping them away from zero energy.
Then Eq.~(\ref{eq:universal_limit}) holds, and one thus concludes that adding any semi-transparent hopping between the boundaries of the HN chain forces the limiting spectral distribution to coincide with the PBC one.

\subsection{Universal aspects of the energy level accumulation}\label{sec:energy_accumulation}

In one-dimensional systems, the spectrum of the infinite-volume operator $\hat H$ is supported on a closed curve, or on a finite union of such curves, in the complex-energy plane. In higher dimensions, by contrast, the spectrum is typically supported on extended two-dimensional regions of $\mathbb C$. For $z$ inside such spectral regions, the corresponding Hermitian double is generically a bulk metal or semimetal. In this case, there is no reason to expect different thermodynamic limits to produce the same limiting Green's function $G_{\Omega}(z,\bar z)$.

\begin{figure}
    \centering

\begin{tikzpicture}[x=0.75pt,y=0.75pt,yscale=-1,xscale=1,scale=0.85]
\draw  [fill={rgb, 255:red, 255; green, 255; blue, 255 }  ,fill opacity=1, line width=.6 ] (204.67,6.17) -- (458.83,6.17) -- (458.83,260.33) -- (204.67,260.33) -- cycle ;
\draw [line width=.6]   (205,131.67) -- (456,131.67) ;
\draw [shift={(458,131.67)}, rotate = 180.2] [color={rgb, 255:red, 0; green, 0; blue, 0 }  ][line width=0.6]    (10.93,-3.29) .. controls (6.95,-1.4) and (3.31,-0.3) .. (0,0) .. controls (3.31,0.3) and (6.95,1.4) .. (10.93,3.29)   ;
\draw  [line width=.6]  (330.66,260.19) -- (330.66,7.84) ;
\draw [shift={(330.66,5.84)}, rotate = 90.28] [color={rgb, 255:red, 0; green, 0; blue, 0 }, line width=.6]    (10.93,-3.29) .. controls (6.95,-1.4) and (3.31,-0.3) .. (0,0) .. controls (3.31,0.3) and (6.95,1.4) .. (10.93,3.29)   ;
\draw  [color={rgb, 255:red, 0; green, 0; blue, 0 }  ,draw opacity=1 ][fill={rgb, 255:red, 56; green, 110; blue, 176 }  ,fill opacity=1.0, line width=.6 ] (260.51,60.99) .. controls (291.86,42.53) and (347,32.27) .. (391,48.27) .. controls (435,64.27) and (451.75,100.66) .. (401.59,171.7) .. controls (351.43,242.75) and (291.86,227.06) .. (260.51,171.7) .. controls (229.16,116.35) and (229.16,79.44) .. (260.51,60.99) -- cycle ;
\draw  [line width=.6]  (420.5,40.27) .. controls (430.35,58.49) and (431.95,65.07) .. (411.45,97.28) ;
\draw [shift={(410.5,98.77)}, rotate = 302.69] [color={rgb, 255:red, 0; green, 0; blue, 0 }  ][line width=0.6]    (10.93,-3.29) .. controls (6.95,-1.4) and (3.31,-0.3) .. (0,0) .. controls (3.31,0.3) and (6.95,1.4) .. (10.93,3.29)   ;
\draw  [color={rgb, 255:red, 0; green, 0; blue, 0 }  ,draw opacity=1 ][fill={rgb, 255:red, 255; green, 255; blue, 255 }  ,fill opacity=1, line width=.6 ] (299.55,77.22) .. controls (307.88,65.67) and (333.36,58.36) .. (356.46,60.9) .. controls (379.57,63.44) and (391.55,74.86) .. (383.22,86.42) .. controls (374.89,97.97) and (349.41,105.28) .. (326.31,102.74) .. controls (303.2,100.21) and (291.22,88.78) .. (299.55,77.22) -- cycle ;
\draw [color={rgb, 255:red, 0; green, 0; blue, 0 }  ,draw opacity=1, line width=.6 ][fill={rgb, 255:red, 215; green, 215; blue, 215 }  ,fill opacity=1 ]   (266,130.48) .. controls (271,94.48) and (308.67,102.71) .. (314.67,138.71) .. controls (320.67,174.71) and (322.67,188.38) .. (306.67,189.71) .. controls (290.67,191.04) and (262,163.48) .. (266,130.48) -- cycle ;
\draw   [line width=.6] (330.66,103.42) -- (330.66,61.84) ;
\draw   [line width=.6] (266,131.67) -- (313.2,131.67) ;
\draw [color={rgb, 255:red, 0; green, 0; blue, 0 }  ,draw opacity=1, line width=.6][fill={rgb, 255:red, 255; green, 255; blue, 255 }  ,fill opacity=1 ]   (344,130.65) .. controls (349,98.38) and (405.5,111.38) .. (394,146.88) .. controls (382.5,182.38) and (361,191.38) .. (350,184.38) .. controls (339,177.38) and (338.5,164.38) .. (344,130.65) -- cycle ;
\draw    [line width=.6] (344,131.67) -- (395,131.67) ;

\draw (336.5,9.4) node [anchor=north west][inner sep=0.75pt][font=\footnotesize, scale=0.9]    {$\Im z$};
\draw (432,139.4) node [anchor=north west][inner sep=0.75pt][font=\footnotesize, scale=0.9]    {$\Re z$};
\draw (210.9,20.73) node [anchor=north west][inner sep=0.75pt]  [font=\footnotesize, scale=0.67]  {$ \begin{array}{l}
\nu _{2n+1} =0\\
G_{\mathrm{PBC}} =G_{\mathrm{OBC}}
\end{array}$};
\draw (328.89,65.74) node [anchor=north west][inner sep=0.75pt]  [font=\footnotesize, scale=0.63]  {$ \begin{array}{l}
\nu _{2n+1} =0\\
G_{\mathrm{B}} =G_{\mathrm{OBC}}
\end{array}$};
\draw (270.45,143.67) node [anchor=north west][inner sep=0.75pt]  [font=\footnotesize, scale=0.7]  {$
\nu _{2n+1} \neq 0$};
\draw (342.4,131.67) node [anchor=north west][inner sep=0.75pt]  [font=\footnotesize, scale=0.63]  {$ \begin{array}{l}
\nu _{2n+1} = 0\\
G_{\mathrm{B}} = G_{\mathrm{OBC}}
\end{array}$};
\draw (392.5,19.4) node [anchor=north west][inner sep=0.75pt]  [font=\footnotesize]  {$\operatorname{spec}(\hat{H})$};

\end{tikzpicture}

\caption{Spectrum of a typical \(d=(2n\,{+}\,1)\)-dimensional non-Hermitian model (equivalently, phase diagram of the corresponding Hermitian double). The resolvent set has four connected components: the far-field component and three holes inside the spectrum (point gaps). One of these point gaps carries a nontrivial winding number (light grey), while the other two and the far-field point gap have \(\nu_{2n+1}=0\) (white). As explained in the text, this implies that in the trivial point gaps and in the far-field component, one has \(G_{\Omega}=G_{\mathrm B}\), and hence the limiting spectral density \(\rho_\Omega(z,\bar z)\) is not supported there.
In the nontrivial component, no analogous conclusion can be drawn immediately, since the corresponding Hermitian double possesses a topological boundary zero mode throughout the entire point gap. Equivalently, the whole component belongs to the spectrum of the non-Hermitian system under semi-infinite boundary conditions (SIBC). Under OBC, energy levels are therefore expected to accumulate throughout this component. The present approach alone, however, does not determine whether the number of accumulating levels is extensive. Consequently, topology by itself does not imply that the limiting spectral density is nonzero there, or that \(G_{\Omega}\neq G_{\mathrm B}\) throughout this component.
}
    \label{fig:phase_diag}
\end{figure}
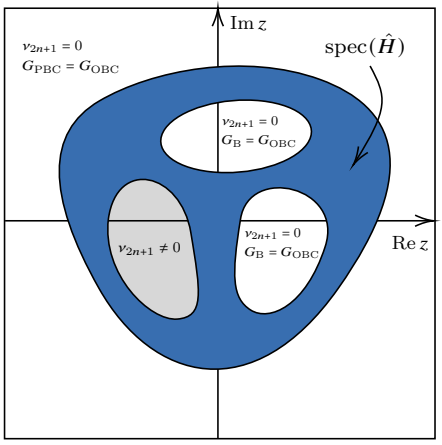

Outside the spectrum of the infinite-volume operator, the situation is more constrained. The spectrum of $\hat H$ partitions the complex plane into connected components of the resolvent set, i.e., into point gaps; equivalently, one may view these components on the Riemann sphere, including the point at infinity. For every $z$ in such a component, the Hermitian double remains gapped and therefore realizes a bulk insulating phase. This phase may be topologically trivial or nontrivial, while the far-field component is necessarily trivial. If the corresponding insulating phase is topologically trivial, Eq.~(\ref{eq:universal_limit}) suggests that the limiting Green's function should be independent of the thermodynamic limit and should coincide with the analytic bulk resolvent,
\[
G_\Omega(z,\bar z)=\bra{i}(z-\hat H)^{-1}\ket{i}. 
\]
Since the right-hand side is analytic throughout the corresponding point gap, the limiting density
\[
\rho_\Omega(z,\bar z)=\frac{1}{\pi}\bar\partial G_\Omega(z,\bar z)
\]
has no support inside such trivial point gaps (see Fig.~\ref{fig:phase_diag}).

This statement should be understood as a stable one. The fact that the bulk phase of the Hermitian double is topologically trivial at a fixed value of $z$ does not, by itself, forbid accidental zero-energy in-gap states localized near the boundary. Such states are not protected by the bulk point-gap topology and can generically be removed by boundary-localized perturbations. In this stable sense, topologically trivial point gaps do not support robust zero-energy boundary modes of the Hermitian double at fixed $z$, which is the content behind Eq.~(\ref{eq:universal_limit}).

There is, however, an important distinction between a fixed-$z$ statement and a statement uniform over an extended set of $z$ values. A boundary perturbation that removes a zero-energy boundary state of $\hat{\mathbb{H}}_{\mathrm{OBC}}(z)$ at one value of $z$ may merely move this zero-energy state to appear in the Hermitian double at a nearby value $\tilde z$. Thus, although zero modes in a trivial point gap are not protected at any prescribed value of $z$, they need not be globally removable as $z$ is varied. In the original non-Hermitian problem, this means that robust spectral sets of boundary origin may occur inside otherwise trivial point gaps, namely those values of $z$ for which the Hermitian double develops zero-energy boundary states. These sets are movable in the $z$-plane by boundary-localized perturbations, but cannot necessarily be eliminated altogether.

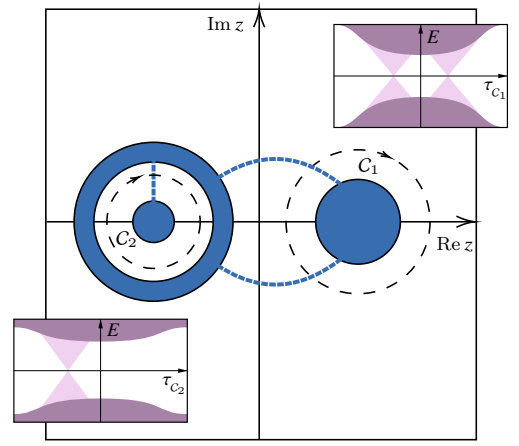
\begin{figure}
    \centering

\begin{tikzpicture}[x=0.75pt,y=0.75pt,yscale=-1,xscale=1,scale=0.85]
\draw  [fill={rgb, 255:red, 255; green, 255; blue, 255 }  ,fill opacity=1, line width=.6 ] (204.67,6.17) -- (458.83,6.17) -- (458.83,260.33) -- (204.67,260.33) -- cycle ;
\draw [line width=.6]   (205,131.67) -- (456,131.67) ;
\draw [shift={(458,131.67)}, rotate = 180.2] [color={rgb, 255:red, 0; green, 0; blue, 0 }  ][line width=0.6]    (10.93,-3.29) .. controls (6.95,-1.4) and (3.31,-0.3) .. (0,0) .. controls (3.31,0.3) and (6.95,1.4) .. (10.93,3.29)   ;
\draw  [line width=.6]  (330.66,260.19) -- (330.66,7.84) ;
\draw [shift={(330.66,5.84)}, rotate = 90.28] [color={rgb, 255:red, 0; green, 0; blue, 0 }, line width=.6]    (10.93,-3.29) .. controls (6.95,-1.4) and (3.31,-0.3) .. (0,0) .. controls (3.31,0.3) and (6.95,1.4) .. (10.93,3.29)   ;
\draw[
  fill={rgb,255:red,56; green,110; blue,176},
  fill opacity=1, line width=.6 
] (268.27,131.67) circle [radius=46.92];
\draw[
  fill={rgb,255:red,56; green,110; blue,176},
  fill opacity=1, line width=.6 
] (389,131.67) circle [radius=25];
\draw[ line width=.6 
] [dash pattern={on 4.5pt off 4.5pt}] 
(389,131.67) circle [radius=42.5];
\draw [color={rgb, 255:red, 56; green, 110; blue, 176 }  ,draw opacity=1 ][dash pattern={on 2.5pt off 0.5pt}, line width=1.5]    (306,106) .. controls (343.35,81.17) and (370.35,103.17) .. (382.35,112.17) ;
\draw [color={rgb, 255:red, 56; green, 110; blue, 176 }  ,draw opacity=1 ][dash pattern={on 2.5pt off 0.5pt}, line width=1.5]    (306,157.34) .. controls (343.35,182.17) and (370.35,160.17) .. (382.35,151.17) ;

\draw[
  fill={rgb,255:red,255; green,255; blue,255},
  fill opacity=1, line width=.6 
] (268.27,131.67) circle [radius=35.23];
\draw[
  fill={rgb,255:red,56; green,110; blue,176},
  fill opacity=1, line width=.6 
] (268.27,131.67) circle [radius=12.23];
\draw [color={rgb, 255:red, 56; green, 110; blue, 176 }  ,draw opacity=1 ][dash pattern={on 2.5pt off 0.5pt},line width=1.5]    (268.27,120.79) -- (268.27,92.79) ;
\draw[
  dash pattern={on 4.5pt off 4.5pt},line width=.6 
] (268.27,131.67) circle [radius=27.54];
\draw [shift={(260,105.1)}, rotate = 154.99] [color={rgb, 255:red, 0; green, 0; blue, 0 }  ][line width=0.6, scale=0.7]    (7.65,-2.3) .. controls (4.86,-0.97) and (2.31,-0.21) .. (0,0) .. controls (2.31,0.21) and (4.86,0.98) .. (7.65,2.3)   ;
\draw [shift={(407.85,93.74)}, rotate = 205.1] [color={rgb, 255:red, 0; green, 0; blue, 0 }  ][line width=0.6, scale=0.7]    (7.65,-2.3) .. controls (4.86,-0.97) and (2.31,-0.21) .. (0,0) .. controls (2.31,0.21) and (4.86,0.98) .. (7.65,2.3)   ;
\draw [line width=.6 ]   (233.04,131.67) -- (256.04,131.67) ;
\draw [line width=.6 ]   (280.5,131.67) -- (303.5,131.67) ;

\draw (298.5,9.4) node [anchor=north west][inner sep=0.75pt][font=\footnotesize, scale=0.9]    {$\Im z$};
\draw (434,139.4) node [anchor=north west][inner sep=0.75pt][font=\footnotesize, scale=0.9]    {$\Re z$};
\draw (387,92.4) node [anchor=north west][inner sep=0.75pt]  [font=\scriptsize]  {$\mathcal{C}_{1}$};
\draw (244,135.4) node [anchor=north west][inner sep=0.75pt]  [font=\scriptsize]  {$\mathcal{C}_{2}$};

\begin{scope}[shift={(10,-44)}, scale=1.6]
\draw[
  fill=white,
  fill opacity=1,
  draw=none
]
(228,37) -- (292,37) -- (292,75) -- (228,75) -- cycle;

\draw[
  fill={rgb,255:red,240; green,210; blue,240},
  fill opacity=1,
  draw opacity=0,
]
(254,37) -- (286,37) -- (270,56) -- cycle;

\draw[
  fill={rgb,255:red,240; green,210; blue,240},
  fill opacity=1,
  draw opacity=0,
]
(254,75) -- (286,75) -- (270,56) -- cycle;

\draw[
  fill={rgb,255:red,240; green,210; blue,240},
  fill opacity=1,
  draw opacity=0,
]
(266,37) -- (234,37) -- (250,56) -- cycle;

\draw[
  fill={rgb,255:red,240; green,210; blue,240},
  fill opacity=1,
  draw opacity=0,
]
(266,75) -- (234,75) -- (250,56) -- cycle;

\draw[
  fill={rgb,255:red,167; green,130; blue,167},
  fill opacity=1,
  draw=none
]
(228,37) -- (292,37)
.. controls (280,37) and (284.8,48.25) .. (260,48.25)
.. controls (235.2,48.25) and (238.4,37) .. (228,37)
-- cycle;

\draw[
  fill={rgb,255:red,167; green,130; blue,167},
  fill opacity=1,
  draw=none
]
(228,75) -- (292,75)
.. controls (280,75) and (284.8,63.75) .. (260,63.75)
.. controls (235.2,63.75) and (238.4,75) .. (228,75)
-- cycle;

\draw[
  draw=black,
  line width=.45
]
(228,37) -- (292,37) -- (292,75) -- (228,75) -- cycle;

\draw[
  draw=black,
  line width=.2
]
(228,56) -- (292,56);
\draw [shift={(292,56)}, rotate = 180.0, scale=0.4] [fill={rgb, 255:red, 0; green, 0; blue, 0 }  ][line width=0.08]  [draw opacity=0] (8.4,-2.1) -- (0,0) -- (8.4,2.1) -- cycle    ;
\draw (282.5,57.5) node [anchor=north west][inner sep=0.75pt]  [font=\small, scale=0.8]  {$\tau_{_{\mathcal{C}_{1}}}$};

\draw[
  draw=black,
  line width=.2
]
(260,37) -- (260,75);
\draw [shift={(260,37)}, rotate = 90.0, scale=0.4] [fill={rgb, 255:red, 0; green, 0; blue, 0 }  ][line width=0.08]  [draw opacity=0] (8.4,-2.1) -- (0,0) -- (8.4,2.1) -- cycle    ;
\draw (261.5,38.5) node [anchor=north west][inner sep=0.75pt]  [font=\small, scale=0.8]  {$E$};
\end{scope}

\begin{scope}[shift={(-179,130)}, scale=1.6]
\draw[
  fill=white,
  fill opacity=1,
  draw=none
]
(228,37) -- (292,37) -- (292,75) -- (228,75) -- cycle;

\draw[
  fill={rgb,255:red,240; green,210; blue,240},
  fill opacity=1,
  draw opacity=0,
]
(262,37) -- (234,37) -- (248,56) -- cycle;

\draw[
  fill={rgb,255:red,240; green,210; blue,240},
  fill opacity=1,
  draw opacity=0,
]
(262,75) -- (234,75) -- (248,56) -- cycle;

\draw[
  fill={rgb,255:red,167; green,130; blue,167},
  fill opacity=1,
  draw=none
]
(228,37) -- (292,37) -- (292,40)
.. controls (280,40) and (284.8,45.25) .. (260,45.25)
.. controls (235.2,45.25) and (238.4,40) .. (228,40) 
-- cycle;

\draw[
  fill={rgb,255:red,167; green,130; blue,167},
  fill opacity=1,
  draw=none
]
(228,75) -- (292,75) -- (292, 72)
.. controls (280,72) and (284.8,66.25) .. (260,66.25)
.. controls (235.2,66.25) and (238.4,72) .. (228,72)
-- cycle;

\draw[
  draw=black,
  line width=.45
]
(228,37) -- (292,37) -- (292,75) -- (228,75) -- cycle;

\draw[
  draw=black,
  line width=.2
]
(228,56) -- (292,56);
\draw [shift={(292,56)}, rotate = 180.0, scale=0.4] [fill={rgb, 255:red, 0; green, 0; blue, 0 }  ][line width=0.08]  [draw opacity=0] (8.4,-2.1) -- (0,0) -- (8.4,2.1) -- cycle    ;
\draw (282.0,57.5) node [anchor=north west][inner sep=0.75pt]  [font=\small, scale=0.8]  {$\tau_{_{\mathcal{C}_{2}}}$};

\draw[
  draw=black,
  line width=.2
]
(260,37) -- (260,75);
\draw [shift={(260,37)}, rotate = 90.0, scale=0.4] [fill={rgb, 255:red, 0; green, 0; blue, 0 }  ][line width=0.08]  [draw opacity=0] (8.4,-2.1) -- (0,0) -- (8.4,2.1) -- cycle    ;
\draw (261.5,38.0) node [anchor=north west][inner sep=0.75pt]  [font=\small, scale=0.8]  {$E$};
\end{scope}

\end{tikzpicture}

\caption{
Schematic spectrum of a generic two-dimensional non-Hermitian system in a semi-infinite geometry. The bulk spectrum, i.e., the spectrum of the infinite-volume operator, is shown as three blue components. The dashed blue lines denote pump-induced boundary spectra inside trivial point gaps. The corresponding pump cycles in the resolvent set are denoted by $\mathcal C_1$ and $\mathcal C_2$. The insets show the spectrum of the Hermitian double of a semi-infinite system as a function of the pump parameters $\tau_{\mathcal C_1}$ and $\tau_{\mathcal C_2}$. A nontrivial pump invariant forces the surface gap of the Hermitian double to close along the cycle, here in the form of boundary Dirac cones. The number of such Dirac cones is fixed by the value of the pump invariant. Consequently, the values of $z$ at which these zero-energy boundary crossings occur belong to the semi-infinite spectrum $\operatorname{spec}(\hat H_{\mathrm{SIBC}})$, and finite-OBC spectra are expected to accumulate on the corresponding lines. Observe that the pump topology along $\mathcal{C}_1$ is related to line-gap topology with respect to the line $\Re z =0$; in contrast, analogous line-gap-topological interpretation is impossible for the pump $\mathcal{C}_2$.
}
    \label{fig:line_gap}
\end{figure}

The origin of this phenomenon is a pump topology of the Hermitian double, which in many cases is inherited from the corresponding nontrivial line-gap topology~\cite{Schindler2023}. 
Even if a connected component of the resolvent set is topologically trivial at every fixed $z$, the component itself may fail to be simply connected. Non-contractible cycles in the $z$-plane can then define nontrivial pumps of the Hermitized Hamiltonian. Interpreting the parameter along such a cycle as an additional synthetic momentum (equivalently, as the time parameter of the cyclic pump), the usual bulk-boundary correspondence implies that the boundary gap of the Hermitian double must close somewhere along the pump, even though the bulk gap remains open throughout the cycle. This mechanism is schematically illustrated in Fig.~\ref{fig:line_gap}.

For concreteness, consider the original non-Hermitian Hamiltonian to be symmetry-free, i.e., in class $\mathrm A$. Hermitization then promotes the problem to class $\mathrm{AIII}$. In spatial dimension $d=2n$, the fixed-$z$ Hermitian double has no stable bulk invariant, and therefore the corresponding point gap carries no stable point-gap topology. However, a pump over a cycle in the resolvent set is effectively $(2n+1)$-dimensional, and class $\mathrm{AIII}$ in odd dimension carries an integer invariant~\cite{Ryu:2010}. In the simplest case, $d=2$, this is the three-dimensional AIII winding number, which counts the number of protected Dirac cones on a clean boundary. Hence, as the pump parameter is varied, the boundary spectrum of the Hermitian double must cross zero energy at a finite set of parameter values. Equivalently, the semi-infinite non-Hermitian spectrum $\operatorname{spec}(\hat H_{\mathrm{SIBC}})$ must contain one-dimensional subsets inside trivial point gaps, in addition to the bulk spectrum. These subsets consist precisely of those values of $z$ for which the Hermitian double $\hat{\mathbb{H}}_{\mathrm{SIBC}}(z)$ has a protected zero-energy boundary state. 
They may be displaced by boundary-localized petrurbations; however, as they are protected by the pump invariant, they cannot be globally removed.
Finite-size OBC spectra are therefore expected to accumulate on such line subsets of trivial point gaps.
This phenomenon includes the formerly investigated formation of chiral edge states in the two-dimensional  non-Hermitian Chern insulator~\cite{Kunst:2018,YaoChern2018}.

At this point, it is important to clarify what can and cannot be concluded from the Hermitization argument. The fact that a subset of the complex plane belongs to the SIBC spectrum does not by itself imply that this subset supports a nonzero limiting OBC eigenvalue \emph{density}. Rather, the presence of a zero-energy boundary mode of the semi-infinite Hermitian double only obstructs the argument based on Eq.~(\ref{eq:universal_limit}) establishing \(G_{\mathrm B}=G_{\mathrm{OBC}}\), but does not automatically imply the opposite. It identifies where OBC eigenvalues may accumulate, but does not guarantee that they do so with finite weight in the normalized thermodynamic limit.

There are two separate reasons for this. First, an eigenvalue \(z\) of a finite OBC Hamiltonian corresponds to an \emph{exact} zero-energy mode of its finite Hermitian double,
\begin{equation}
	z\in\operatorname{spec}(\hat H_{\mathrm{OBC},N})\quad\Leftrightarrow\quad0\in\operatorname{spec}[\hat{\mathbb{H}}_{\mathrm{OBC},N}(z)].
\end{equation}
The topological zero modes of the semi-infinite double need not remain exact in a finite system: modes localized at opposite boundaries may hybridize, turning the zero-energy crossing into an avoided crossing. Thus, even when every \(z\) in a bulk point gap belongs to the SIBC spectrum, the exact zero-energy crossings of the finite Hermitian double may occur only on a smaller subset. The unidirectional Hatano-Nelson model discussed in Sec.~\ref{sec:limsup} [cf.~Fig.~\ref{fig:src_convergence_and_corner_modes}] provides a simple example. The entire disk \(|z|<1\) supports a topological zero mode of the semi-infinite Hermitian double and therefore belongs to the SIBC spectrum, whereas for every finite OBC truncation an exact zero-energy crossing occurs only at \(z=0\).

Second, even when exact crossings occur and produce OBC eigenvalues, their total algebraic multiplicity need not scale extensively with the system volume, as required to contribute a finite weight to the limiting normalized density. The Hermitization argument alone does not automatically determine this scaling. For example, the nontrivial point gap in Fig.~\ref{fig:phase_diag} is expected to support boundary-localized states.\footnote{Throughout a point gap, the bulk spectrum of the Hermitian double is gapped at zero energy. Hence, its only stable zero-energy crossings are those of topological boundary modes. At an exact finite-OBC crossing \(z\), let\[\hat{\mathbb{H}}_{\mathrm{OBC},N}(z)\Psi_N=0,\qquad\Psi_N=(u_N,v_N)^T.\]Since \(\Psi_N\) is a boundary mode of a gapped Hermitian topological system, its spatial profile and localization are determined by the usual methods for Hermitian topological boundary states and are, e.g.,  captured by the corresponding semi-infinite Hermitized problem. Chiral projection onto the negative-chirality sector gives \((0,v_N)^T\), for which\[(H_{\mathrm{OBC},N}-z)v_N=0.\] Thus, \(v_N\) is the corresponding right eigenvector of the original non-Hermitian Hamiltonian, and its spatial localization follows unambiguously from that of the Hermitian boundary mode \(\Psi_N\). For example, in one dimension, this construction recovers the usual localization rules for the non-Hermitian skin effect~\cite{Kawabata:2019}.}
In one dimension, the associated OBC eigenvalues may occur with extensive total multiplicity, producing the usual non-Hermitian skin effect. By contrast, boundary states arising solely from a strong three-dimensional point-gap winding invariant are expected to scale with the boundary area and are therefore subextensive~\cite{Denner_2021}. They remain visible in the limiting spectrum viewed as a set, but disappear from the normalized limiting density.

In fact, under the assumptions of the amoeba theory~\cite{Wang:2024}, only point gaps carrying nontrivial weak one-dimensional winding invariants can support an extensive spectral weight. In this sense, the extensive point-gap skin effect is governed by an essentially one-dimensional topological mechanism, whereas strong higher-dimensional point-gap invariants generically produce subextensive boundary spectral features. The present Hermitization approach captures where both types of boundary spectra may occur, but does not by itself determine whether their algebraic multiplicity is extensive.

Analogously, the pump-induced spectral sets do not contribute to the limiting density of states. For instance, in two dimensions, the above mechanism may produce line-like accumulation sets in the complex-energy plane, and one might therefore wonder whether such lines support nonzero line charges in the limiting spectral distribution.
However, away from the line itself, the point gap is still trivial, and the Hermitian double has no stable zero-energy boundary modes. Therefore, on both sides of a line charge distribution, the limiting Green's function agrees with the analytic bulk resolvent,
\begin{equation}
G_{\mathrm{OBC}}(z,\bar z)=G(z).
\end{equation}
As such, there is no jump of the ``electric field'' $G_{\mathrm{OBC}}$ across the line charge, meaning its density vanishes identically, in agreement with the conclusion of the amoeba theory. Thus, the pump-induced boundary levels, although visible in the open- or semi-infinite-boundary spectrum as a set, carry zero weight in the thermodynamic-limit density of states.

In higher dimensions, the geometry of these pump-induced spectral sets can be more complicated. The low-energy boundary theory associated with a nontrivial pump need not consist only of isolated Dirac cones; generically, the zero-energy boundary spectrum may form nodal structures.\footnote{
To see a concrete example, consider the Hermitian double (Altland-Zirnbauer class AIII) of a generic non-Hermitian Hamiltonian in 4D. Upon introducing the pump parameter, this is extended into a Hamiltonian in 5D.
Such a system is characterized by a $\mathbb{Z}$ winding number~\cite{Ryu:2010}. 
Taking the $4\times 4$ Dirac matrix $\mathbb{\Gamma}_5$ to denote the chiral operator, one might expect the topological surface states to be captured by the Dirac Hamiltonian ${\propto}\sum_{a=1}^4 k_a \mathbb{\Gamma}_a$ where $
\{k_a\}_{a=1}^4$ are momenta parallel with the 4D surface. 
However, the point-like band degeneracy of this surface Hamiltonian inside its 4D surface Brillouin zone (SBZ) contradicts that band degeneracies in class AIII are generically of codimension $2$ in the parameter space~\cite{Bzdusek:2017}. 
Indeed, one can enrich the surface Hamiltonian with $k$-independent terms proportional to $\mathbb{\Gamma}_{15}$, $\mathbb{\Gamma}_{25}$, $\mathbb{\Gamma}_{35}$, and $\mathbb{\Gamma}_{45}$. 
The resulting Hamiltonian is analytically diagonalizable~\cite{Iliasov:2026}, with the perturbations generically inflating the Dirac point into a nodal manifold of codimension $2$ (i.e., a 2D surface inside the 4D SBZ).
} Consequently, the set of pump parameters for which the boundary gap closes need not be discrete, and $\operatorname{spec}(\hat H_{\mathrm{SIBC}})$ may contain extended volume-like subsets lying inside otherwise trivial point gaps, in addition to bulk spectrum and nontrivial point gaps. 

We therefore arrive at the following picture. Point-gap topology determines whether robust boundary-localized states may occur at fixed \(z\). In particular, nontrivial weak one-dimensional point-gap invariants can generate an extensive number of such states and thereby contribute to the limiting normalized spectral density. Pump topology, on the other hand, controls a more refined property of the spectrum as a set: it can force protected boundary spectral subsets even inside point gaps that are locally trivial at fixed $z$. These subsets are movable by boundary-localized perturbations but cannot be globally removed. Their total number remains subextensive, however, and they therefore carry no weight in the limiting normalized density of states.

As such, at the level of limiting densities $\rho_\Omega(z,\bar z)$, the support is expected to be governed only by the bulk spectrum together with (possibly) the nontrivial point gaps. Since the far-field point gap is always trivial, the extent of the support of $\rho_\Omega(z,\bar z)$ is expected to be smaller or equal than that of PBC (Brown) DOS $\rho_{\mathrm B}(z,\bar z)$ support, which in typical cases occupies the full spectral region and therefore has maximal extent. In other words, the limiting PBC spectral distribution (the Brown measure of $\hat{H}$) is expected to have the largest spectral radius~\cite{Wang:2024}.

Importantly, in higher dimensions, the collapse of limiting eigenvalue densities is not tied uniquely to nontrivial point-gap topology. Even in the absence of such topology, the limiting density may reorganize within the extended PBC spectral support, without changing the multipole moments that constrain the corresponding Green's functions. The special role of point-gap topology in the collapse of spectral measures becomes especially sharp in 1D. Namely, in the case of a one-dimensional problem with trivial point gaps, the spectral density cannot freely reorganize on the one-dimensional PBC spectrum: it is fixed by the boundary values of the corresponding thermodynamic Green's functions on the adjacent point gaps. Hence, nontrivial point-gap topology is effectively the only mechanism for spectral collapse in one dimension.

\subsection{Symmetry-protected spectra}\label{sec:sym-protected}

We now briefly discuss how the preceding picture is modified in the presence of symmetry, namely when the original non-Hermitian system is constrained to belong to one of the inherent non-Hermitian symmetry classes~\cite{Kawabata:2019}. 
This discussion is not intended to be an exhaustive characterization of all 38 non-Hermitian symemtry classes; rather, it aims to illustrate the key modifications of the spectral features that can potentially take place.

From the perspective of Hermitization, the essential difference is that the Hermitized Hamiltonian $\hat{\mathbb{H}}(z)$ possesses, in addition to the intrinsic chiral symmetry
\(
	\Sigma=\tau_z\otimes 1,
\)
further symmetries inherited from the original non-Hermitian Hamiltonian. Here, $\tau_i$ denote Pauli matrices acting in the Hermitization space.

As an example, suppose that the original non-Hermitian Hamiltonian $h(k)$ itself possesses a sublattice symmetry $\mathcal{S}$ (see, e.g., Appendix~\ref{a:NH_SSH}); namely $\mathcal{S}h(k)\mathcal{S}^{-1}=-h(k)$. The corresponding lifted operator
\begin{equation}
	\tilde S=\tau_0\otimes \mathcal S
\end{equation}
introduces the following action on the Hermitized Hamiltonian:
\begin{equation}\label{eq:nh-sublat}
	\tilde S\hat{\mathbb{H}}(k,z)\tilde S^{-1}=-\hat{\mathbb{H}}(k,-z).
\end{equation}
In contrast to the intrinsic Hermitization-induced chiral symmetry $\Sigma$, this symmetry does not act locally in the spectral parameter, but instead maps $z\mapsto-z$.

Similarly, non-Hermitian chiral symmetry $\Gamma$~\cite{Kawabata:2019}, which constrains the non-Hermitian Hamiltonian through $\Gamma h^\dagger(k)\Gamma^{-1}=-h(k)$, can be lifted by introducing
\(
	\tilde\Gamma=\tau_x\otimes\Gamma,
\)
	which satisfies
\begin{equation}\label{eq:nh-chiral}
	\tilde\Gamma\hat{\mathbb{H}}(k,z)\tilde\Gamma^{-1}=-\hat{\mathbb{H}}(k,-\bar z).
\end{equation}

Proceeding in the same manner, one can lift all elementary non-Hermitian symmetry constraints~\cite{Denner_2023}. Together with their mutual algebra and the intrinsic symmetry $\Sigma$, these lifts reproduce the Hermitian representatives associated with the 38 non-Hermitian symmetry classes~\cite{Kawabata:2019}.

The important feature of these lifted symmetries is their action on the spectral parameter. In general, they map $z$ to one of \(z, \bar z, - z, -\bar z\). The maps $z\mapsto\bar z$ and $z\mapsto-\bar z$ are reflections across the real and imaginary axes, respectively, while $z\mapsto-z$ is inversion about the origin. Correspondingly, the fixed locus of a lifted symmetry may be the entire complex plane, the real axis, the imaginary axis, or only the origin.
 
First we discuss the case where the lifted symmetry acts locally in $z$; this is the case for, e.g., $\mathrm{TRS}^\dagger$, \[\mathcal{C}_+ h^T(k)\mathcal{C}_+^{-1}=h(-k).\] 
This symmetry is lifted to $\tilde{C}_+=\left(\tau_x\otimes \mathcal{C}_+\right)K$, with $K$ being the complex conjugation, ensuring:
 \begin{equation}
 	\tilde{C}_+ \hat{\mathbb{H}}(k,z)\tilde{C}_+^{-1}=\hat{\mathbb{H}}(-k,z).
 \end{equation}
At every $z$, $\tilde{C}_+$ acts like a time-reversal symmetry (TRS) on the Hermitian double. Moreover, it anticommutes with the intrinsic chiral symmetry $\Sigma$, dictating that the auxiliary Hermitian system does not belong to the complex class $\mathrm{AIII}$, but rather to one of the real symmetry classes  $\mathrm{DIII}$ or $\mathrm{CI}$, depending on the sign of $\mathcal{C}_+\mathcal{C}_+^*$. 
As such, both the point-gap and the pump invariants are governed by the corresponding real classes of the tenfold way characterization~\cite{Ryu:2010}, leading, through the bulk--boundary correspondence, to spectral phenomena analogous to those discussed in the previous section.

A slightly more subtle situation arises when the lifted symmetry does not fix a generic value of $z$. In this case, an individual Hermitized Hamiltonian $\hat{\mathbb{H}}(z)$ generically possesses only the intrinsic chiral symmetry $\Sigma$. The lifted symmetry instead relates $\hat{\mathbb{H}}(z)$ to another member $\hat{\mathbb{H}}({f(z)})$ of the Hermitized family. On the fixed locus $f(z)=z$, however, it becomes an additional local symmetry and refines the symmetry class of the Hermitian double.

For example, non-Hermitian chiral symmetry, Eq.~(\ref{eq:nh-chiral}), maps $z\mapsto-\bar z$ and therefore acts locally on the imaginary axis, $z\in i\mathbb R$, whereas sublattice symmetry, Eq~(\ref{eq:nh-sublat}), maps $z\mapsto-z$ and acts locally only at $z=0$. More generally, the symmetry-refined structure is confined to the corresponding fixed loci: the real axis, the imaginary axis, or the origin.

On each such locus, one can again apply the logic developed in the preceding subsection, now using the refined symmetry class of $\hat{\mathbb{H}}(z)$. As a concrete example, consider a non-Hermitian Hamiltonian with ordinary time-reversal symmetry,
\begin{equation}
	\mathcal T_+ h^*(k)\mathcal {T}^{-1}_+ = h(-k).
\end{equation}
Its lift, $\tilde{T}_+=\left(\tau_0 \otimes \mathcal{T}_+\right)K$,  satisfies:
\begin{equation}\label{eq:nh-trs}
	\tilde{T}_+\hat{\mathbb{H}}(k,z)\tilde{T}_+^{-1}=\hat{\mathbb{H}}(-k,\bar z).
\end{equation}
It therefore reduces to an ordinary Hermitian time-reversal symmetry on the real axis, $z\in\mathbb R$. Since $\tilde{T}_+$ commutes with $\Sigma$, the Hermitized Hamiltonian belongs on this axis to the class $\mathrm{BDI}$ or $\mathrm{CII}$, depending on the sign of $\mathcal{T}_+\mathcal{T}_+^*$.

As $z$ is varied along the real axis, $\hat{\mathbb{H}}(z)$ is either bulk-gapped, when $z$ lies in the intersection of the bulk resolvent set with the real axis, or bulk-gapless, when $z$ intersects the bulk spectrum. The real axis can therefore be interpreted as a one-parameter topological phase diagram of Hermitian Hamiltonians in class $\mathrm{BDI}$ or $\mathrm{CII}$, in the same way that the entire complex spectral plane played the role of a class-$\mathrm{AIII}$ phase diagram in the symmetryless case. Each connected gapped interval can be assigned the corresponding $\mathrm{BDI}$ or $\mathrm{CII}$ bulk invariant, while the far-field interval is necessarily topologically trivial. Consequently, in analogy with 
the symmetryless case, the bulk-boundary correspondence ensures that if a gapped interval of $\hat{\mathbb{H}}(z)$ carries a nontrivial bulk invariant, then this interval belongs to the SIBC spectrum of the original non-Hermitian problem, while the OBC spectrum is expected to accumulate along it. This reasoning, however, applies only on the symmetry-fixed locus and does not, in general, protect spectral features away from it.

A generic non-Hermitian symmetry class may contain several symmetry constraints simultaneously, so that the real axis, the imaginary axis, and the origin may be governed by distinct symmetry-refined Hermitian classes. The corresponding spectral collapse is therefore confined to the nontrivial components of the resolvent set restricted to these fixed loci, thereby providing a boundary realization of the point-gap invariants of the 38 non-Hermitian symmetry classes~\cite{Kawabata:2019,Denner_2023}, generalizing the role played by the winding number in the symmetryless case.

The limiting case of this construction occurs when the fixed locus consists only of the origin. This happens, for example, in a system possessing only sublattice symmetry, Eq~(\ref{eq:nh-sublat}), such as the non-Hermitian Su-Schrieffer-Heeger (nH-SSH) model~\cite{Yao:2018,Kunst:2018} discussed in Appendix~\ref{a:NH_SSH}. In this case, there is no extended symmetry-refined parameter space and hence no nontrivial phase-diagram interpretation involving different connected components or a trivial far-field region. Nevertheless, if $\hat{\mathbb{H}}(z\,{=}\,0)$ is bulk-gapped, the corresponding refined invariant can diagnose a symmetry-protected zero-energy boundary state at the origin. We include a dedicated discussion of the non-trivial topology of the nH-SSH model in Appendix~\ref{a:NH_SSH}.

The pump construction becomes more intricate in the presence of symmetry. In particular, from the perspective of boundary spectral flow, symmetry-protected states associated with the point-gap invariants discussed above may contribute to the pump response. Consequently, the resulting pump invariant need not describe topology intrinsic to the additional pump dimension, but may contain a contribution inherited from a lower-dimensional symmetry-refined point-gap \mbox{invariant}. 
We do not attempt here a systematic classification of such pumps, nor do we discuss the recovery of line-gap invariants from them, as was done in the symmetryless case. 
Nevertheless, we wish to mention here several pump-inspired observations that may be useful for further studies. 

As far as we can see, these observations do not directly translate into corresponding line-gap phenomena, owing to the limitations of the line-gap construction in the presence of symmetry. 
Recall that our notion of the pump relies essentially on the resolvent set in the parameter space not being simply-connected.
A noncontractible pump path need not be deformable into one passing through the point at infinity, as illustrated by the path $\mathcal C_2$ in Fig.~\ref{fig:line_gap}. 
In particular, a bounded component of the resolvent set may itself be non-simply-connected and therefore support a nontrivial pump, even though no corresponding line-gap exists.

As an example, consider a system with sublattice symmetry and an inversion-invariant loop contained in a bounded component of the resolvent set. Parametrizing the loop by an auxiliary momentum $q$ such that
\begin{equation}
	z(q+\pi)=-z(q),
\end{equation}
the lifted sublattice symmetry in Eq.~\eqref{eq:nh-sublat} acts on the auxiliary Hermitian system as
\begin{equation}
	\tilde S\hat{\mathbb{H}}(k,q)\tilde S^{-1}=-\hat{\mathbb{H}}(k,q+\pi).
\end{equation}
Combining $\tilde S$ with the intrinsic chiral symmetry $\Sigma$, one obtains
\begin{equation}
	U\hat{\mathbb{H}}(k,q)U^{-1}=\hat{\mathbb{H}}(k,q+\pi),\qquad U=\Sigma\tilde S.
\end{equation}
Thus, in addition to its intrinsic class-$\mathrm{AIII}$ symmetry, the $(d+1)$-dimensional auxiliary family possesses a momentum-shift symmetry $q\mapsto q+\pi$. Its real-space implementation requires an additional gauge structure and is therefore naturally related to projective crystalline (or momentum-space-nonsymmorphic) symmetries~\cite{Zhao:2020,Chen:2022,Chen:2023,Konig:2026}. 
Whether this structure supports additional pump invariants, and which spectral features such invariants enforce in the SIBC/OBC geometries, remains an interesting question for further study.

A related question is whether the origin must lie outside this resolvent component. 
If the component contains the origin, 
the inversion-invariant loop may
be continuously contracted to the constant path 
at $z=0$ without closing the bulk gap or breaking the symmetry. 
In the symmetryless case, such a contraction would imply the absence of an invariant intrinsic to the additional pump dimension. 
Here, however, the situation is less immediate: at $z=0$, the Hermitized Hamiltonian possesses both the intrinsic chiral symmetry $\Sigma$ and the lifted sublattice symmetry $\tilde S$, and may therefore carry a nontrivial $d$-dimensional $\mathrm{AIII}+\mathrm{AIII}$ invariant.\footnote{The formation of a pair of $\mathrm{AIII}$ invariants in systems with sublattice symmetry at $z=0$ is also encountered and further clarified in the context of the non-Hermitian SSH model in Appendix~\ref{a:NH_SSH}.}  
This suggests that even a contractible pump may retain an inherently lower-dimensional topology inherited from the symmetry-refined point-gap invariant at the origin, potentially giving rise, through the bulk-boundary correspondence, to additional pump-induced spectral features visible in the SIBC/OBC geometries.
We leave a dedicated analysis of such prospective equivariant pump topology in non-Hermitian Hamiltonians open to future studies.

\section{Conclusion}\label{sec:conclusion}

In this work, we clarified the apparent ambiguity in defining the bulk spectral function of a non-Hermitian lattice problem. 
We showed that this DOS ambiguity is representational rather than fundamental from the perspective of bulk dynamics far from boundaries. 
The universal information about the bulk evolution is encoded not in the limiting eigenvalue densities themselves, but in the far-field frequency values of the Green’s functions, whose thermodynamic limits always agree for/in finite-range tight-binding Hamiltonians.
Equivalently, these far-field Green’s functions are determined only by the multipole moments of the limiting densities. 
The insensitivity of bulk dynamics is reflected in the fact that all limiting densities share identical multipole moments, regardless of the thermodynamic limiting procedure used to construct the infinite lattice. 
However, mixed moments [Eq.~(\ref{eqn:mixed-moments})] are not fixed by evolution alone, so non-Hermitian extended systems may generally admit different limiting spectral densities. 
This contrasts sharply with Hermitian systems, where the reality of the spectrum removes the freedom in transverse moments, implying that their far-field Green’s function uniquely determines the limiting density.
	
As such, even in the non-Hermitian setting, limiting spectral densities remain related, though not necessarily directly, to local observable bulk phenomena. It is therefore desirable to construct a physically consistent spectral function directly at the level of the infinite operator, without referring to a particular approximating sequence. Such an approach would parallel the standard Hermitian notion of a spectral measure, which plays this role and is definable for infinite operators independently of the limiting procedures. We emphasize that such a tool, usually called the Brown measure, is indeed available for a broad class of non-Hermitian operators. We argue that, for translationally-invariant Euclidean lattice systems, this density coincides with the limiting naive eigenvalue distribution of PBC clusters. However, the construction itself is substantially more general and naturally extends beyond flat lattices, including non-Hermitian Hamiltonians on hyperbolic lattices~\cite{Shen:2025,Hu:2025}, trees~\cite{Sun:2024,Hamanaka:2025,Hatano:2026}, and related structures.
	
The usefulness of the Brown measure is twofold. First, it is supported on the spectrum of the infinite operator, which is why it provides a natural non-Hermitian analogue of a spectral measure. Second, the Hermitization procedure used to define it gives a systematic way to determine where eigenvalues of finite truncations with specific boundary conditions may accumulate relative to the intrinsic spectral support of the infinite system. In fact, it identifies where the limiting Green's functions necessarily coincide with the Brown Green's function, enabling standard tools from complex analysis to constrain not only where, but also how, eigenvalues of approximating sequences accumulate in the complex plane.
	
Within this framework, Hermitization constrains where a limiting spectral density may differ from the Brown measure, but does not by itself guarantee such a difference. In every topologically trivial point gap, the limiting Green's function necessarily coincides with the Brown Green's function; through the two-dimensional Gauss law, this excludes any finite limiting spectral weight from these regions. A nontrivial point gap removes this obstruction but does not automatically produce an extensive spectral contribution. Under the assumptions of the amoeba theory~\cite{Wang:2024}, only point gaps carrying nontrivial weak one-dimensional winding invariants can support such an extensive weight. At the same time, Hermitization shows that strong higher-dimensional point-gap invariants can enforce robust families of boundary states whose number does not scale with the system size; although subextensive in the normalized density, these remain important features of the OBC/SIBC spectrum. 
It is thus crucial that the normalized spectral density is distinguished from the spectrum as a set.

Pump topology of the Hermitian double, possibly (but not necessarily) inherited from line-gap topology, may force boundary-induced spectral subsets inside otherwise trivial point gaps. These subsets are visible in semi-infinite and open geometries. Importantly, they may be moved by boundary perturbations but not globally removed. Being boundary-induced and subextensive, however, they are not expected to contribute to the thermodynamic-limit spectral density.

Several qualifications are in order. Our arguments apply to bounded tight-binding Hamiltonians and to approximating sequences converging strongly to the infinite operator. They concern thermodynamic-limit spectral distributions and Green's functions relevant to bulk observables, rather than arbitrary finite-size spectra. Nevertheless, the Hermitization procedure itself appears to define, at finite values of the Hermitization regulator, regularized spectral distributions that may approximate the Brown measure under arbitrary boundary conditions. The sensitivity on this regulator is reminiscent of the role of the pseudospectral regulator~\cite{Okuma:2020b,Trefethen2005}, suggesting an interesting, though not direct, relation between the two constructions.
	
Extension of the presented ideas, rooted in the invariance of the bulk dynamic off the boundary condition, to genuinely Lindbladian dynamics, continuum systems, or interacting many-body settings requires additional considerations beyond the specific mathematical context of our work. 
    
Furthermore, noting that boundary spectral accumulation tied to nontrivial stable point-gap topology is, by the presented analysis, related to the appearance of zero-energy modes in the Hermitian doubled Hamiltonian, it will be interesting to consider whether analogous collapse occurs for points gaps that are nontrivial in refined classifications. Notable examples to consider include higher-order topology protected by crystalline symmetry~\cite{Liu:2019,Okugawa:2020,Zhang:2021} and weak topology protected by translation symmetry~\cite{Okuma:2020,Hofmann:2020}. A paradigmatic setting to investigate these question is provided by two-dimensional models, where point gaps carry no stable topological invariant.

\begin{acknowledgments}
We would like to thank Evgen Len, Wei-Jie Chan, Kyle Monkman, Zolt\'{a}n Guba, Dan Mao, Shu Hamanaka, and Tsuneya Yoshida for valuable discussions.
M.~P., A.~I., and T.~B.~were supported by the Starting Grant No.~211310 by the Swiss National Science Foundation (SNSF). 
A.I. acknowledges support from the UZH Postdoc Grant No.~FK-24-104. E.J.B. is supported by the Knut and Alice Wallenberg Foundation (2023.0256) and the Göran Gustafsson Foundation for Research in Natural Sciences and Medicine. 
\end{acknowledgments}

\appendix

\section{Direct evaluation of $G_{\mathrm{PBC}}(z,\bar z)$ \texorpdfstring{\nopagebreak\\}{} in the Hatano-Nelson model.} \label{a:evaluation_of_the_PBC_GF}

Here, we evaluate the Green's function of the PBC Hatano-Nelson model. It is given by the following expression:
\begin{equation}\label{eq:general_HN_starting_form_app}
    G_{\mathrm{PBC}}(z,\bar z) = \frac{1}{2\pi}\int_0^{2\pi} \mathrm d k \frac{1}{z-t_Re^{ik}-t_L e^{-ik}}
\end{equation}
This integral can be reduced to the contour integral form:
\begin{equation}\label{eq:general_HN_contour_form}
    G_{\mathrm{PBC}}(z,\bar z)=-\frac{1}{2\pi i} \oint_C\frac{\mathrm{d}\omega}{t_R\omega^2-z\omega+t_L}
\end{equation}
with $C$ being the unit-circle counterclockwise contour.

The polynomial in the denominator can be represented as
$$t_R\omega^2-z\omega+t_L=t_R\left(\omega-\frac{\omega_+(z)}{2t_R}\right)\left(\omega-\frac{\omega_-(z)}{2t_R}\right),$$ where 
\begin{equation}
\label{eqn:omega-roots-vs-z}
\omega_\pm(z)={z\pm\sqrt{z^2-4t_Rt_L}}.
\end{equation}
Assume $t_L\geq t_R$, and denote $a=t_L+t_R$, $b=t_L-t_R$ while introducing %
$c=\sqrt{a^2-b^2}=2\sqrt{t_Rt_L}$. 
Note that $a,b$ and $c$ represent, correspondingly, the big and the small semiaxes, and the linear eccentricity of the ellipse in the complex plane on which the integrand in Eq.~(\ref{eq:general_HN_starting_form}) is singular.

By Vieta's formula, $\omega_+(z)\omega_-(z)=c^2$, thus $|\omega_-(z)|=c^2/|\omega_+(z)|$, i.e., $|\omega_+(z)|$ has the same configuration of contour lines as $|\omega_-(z)|$.
Contour lines of $|\omega_+(z)|$ are located on ellipses: $|z+\sqrt{z^2-c^2}|=r\ \Rightarrow z=\frac{1}{2}(r e^{i\phi}+\frac{c^2}{r}e^{-i\phi})$. Those ellipses have big and small semi-axes, and linear eccentricity: 
\begin{equation}
    \tilde a=\frac{1}{2}\left( r +\frac{c^2}{r} \right),\ \tilde b = \frac{1}{2}\left( r -\frac{c^2}{r} \right),\ \tilde c = \sqrt{\tilde a^2-\tilde b^2}=c.
\end{equation}
Note that $|\omega_+(z)|>c$, while $|\omega_-(z)|=c^2/|\omega_+(z)|<c$ on the same ellipse.

We are interested in the conditions under which either $|\omega_+(z)|<2t_R=a-b$ or $|\omega_-(z)|<2t_R=a-b$, which correspond to the occurrence of poles of the integrated function in Eq.~(\ref{eq:general_HN_contour_form}) inside the unit-circle contour $C$. Note that $a-b\leq c$; $|\omega_+(z)|$ does not satisfy this condition  for all $z$ while $|\omega_-(z)|<2t_R$ gives the following geometric set of points: first, $|\omega_-(z)|=2t_R=a-b$ gives an ellipse with $\tilde a = a,\ \tilde b = b,\ \tilde c = c$. Then, $|\omega_-(z)|<2t_R$ is satisfied outside that ellipse.

It is now possible to find the value of the integral in Eq.~(\ref{eq:general_HN_contour_form}) using the residue theorem. 
One has $G_{\mathrm{PBC}}(z,\bar z)=0$ for all $z$ inside the PBC spectral ellipse. Outside of it,  $G_{\mathrm{PBC}}(z,\bar z)$ is equal to $(z^2-4t_Rt_L)^{-{1}/{2}}$:
\begin{equation}
    G_{\mathrm{PBC}}(z,\bar z)=
    \begin{cases}
            \frac
    {
        1
    }
    {
        \sqrt{z^2-4t_Rt_L}
    },& |z-c|+|z+c|\geq2a \\
    0,& \text{otherwise.}\\
    \end{cases}
\end{equation}

\section{Electrostatic potentials and topological \texorpdfstring{\nopagebreak\\}{} boundary states of non-Hermitian SSH model}\label{a:NH_SSH}

\begin{figure}[!t]
\includegraphics[width=1\linewidth]{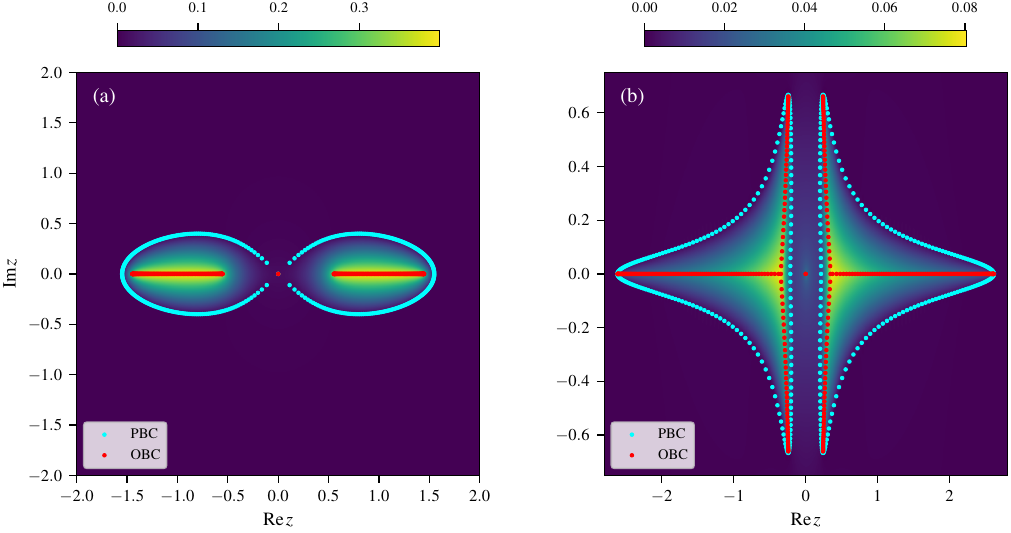}
\caption{The difference $\Phi_{\mathrm{OBC}}-\Phi_{\mathrm{PBC}}$ [cf.~Fig.~\ref{fig:bidirectional_1D_HN_model}] as a function of $z$ for the non-Hermitian SSH model, Eq.~(\ref{eq:NH_SSH}), for various choices of parameters of the model: (a)~For $t_1=0.6, t_2=1, t_3=0$ and $\gamma=-4/5$.
(b)~For $t_1=1, t_2=1, t_3=0.7$ and $\gamma=4/3$.
In both panels, cyan (red) dots indicate the PBC (OBC) spectrum of a chain with $L=300$ unit cells.
}
\label{fig:NH_SSH_Potentials}
\end{figure}

To demonstrate further the methodology developed in the main text, we consider here the non-Hermitian Su-Schrieffer-Heeger (nH-SSH)~\cite{Yao:2018,Kunst:2018} model with the Bloch Hamiltonian:
\begin{align}\nonumber
    h(e^{ik})
    =&
    \left[
        t_1+(t_2+t_3)\cos k
    \right]\sigma_x
    \\
    &+\label{eq:NH_SSH}
    \left[
        (t_2-t_3)\sin k
        +
        i\frac{\gamma}{2}
    \right]\sigma_y,
\end{align}
where $\sigma_{x,y,z}$ are Pauli matrices. We consider the OBC and PBC potentials for this model for two different choices of parameters as specified in caption to Fig.~\ref{fig:NH_SSH_Potentials}.

For this model, the corresponding distributions of eigenenergies for chains of length $L=300$ are shown in Fig.~\ref{fig:NH_SSH_Potentials}, for various choices of parameters. In addition, we plot the difference between the potentials,
\[
\Phi_{\mathrm{OBC}}(z,\bar z)-\Phi_{\mathrm{PBC}}(z,\bar z),
\]
alongside their respective eigenvalue densities.

The plot suggests that, indeed, the differences between potentials are non-constant (equivalently, the difference of the respective Green's functions are non-zero) only inside the nontrivial point gaps. Correspondingly, the OBC spectra accumulate inside the nontrivial point gaps (corresponding to the inside of the PBC spectra shown in cyan) as argued in Sec.~\ref{sec:limsup}.

The point gaps, together with their boundaries, are naturally identified with the spectrum of the corresponding system in the semi-infinite boundary condition (SIBC) geometry. Indeed, inside the point gaps, the Hermitian double of the SIBC system, Eq.~(\ref{eq:Hermitian_Doubled_Hamiltonian}), has a topological zero mode in its spectrum and is therefore not invertible. Thus, the OBC spectra accumulate inside the SIBC spectra.

However, as clearly seen in Fig.~\ref{fig:NH_SSH_Potentials}(b), the OBC spectrum additionally contains the eigenvalue \(z=0\). Its appearance may seem surprising because the origin lies in an otherwise trivial point gap---in fact, in the far-field point gap.

The origin of this state can be understood from the point of view of Sec.~\ref{sec:sym-protected}. In fact, the model, Eq.~(\ref{eq:NH_SSH}), possesses additional non-Hermitian sublattice symmetry, $\mathcal S=\sigma_z$ [with $\sigma_i$ being Pauli matrices acting in the basis of Eq.~(\ref{eq:NH_SSH})]. 
This symmetry is lifted to the Hermitian double as
\[
\tilde S = \tau_0 \otimes \sigma_z. 
\]
The action of the lifted symmetry on the Hermitian double is given by
\[\tilde S\,\hat{\mathbb{H}}(k,z)\,\tilde S^{-1}
=-\hat{\mathbb{H}}(k,-z),\]
cf.~Eq.~(\ref{eq:nh-sublat}).

Therefore, the analysis presented in Sec.~\ref{sec:sym-protected} suggests that the origin is the fixed locus of $\tilde S$, and thus there the Hermitized operator carries a refined invariant. Indeed, at the origin, $\tilde S$ acts like an additional chiral symmetry. Thus, \(\hat{\mathbb{H}}(z=0)\equiv \hat{\mathbb{H}}_0\) possesses two commuting chiral symmetries, namely lifted $\tilde S$ and intrinsic Hermitization chiral symmetry, \(\Sigma=\tau_z\otimes\sigma_0\). Their product,
\[
U=\Sigma\tilde S=\tau_z\otimes\sigma_z,
\]
commutes with \(\mathbb H_0\), allowing the Hermitian double to be decomposed into two sectors, characterized respectively by \(U=\pm1\). Each sector remains in class $\mathrm{AIII}$ and may carry its own one-dimensional winding number, \(\nu_+\) and \(\nu_-\). 
Their sum gives the ordinary $\mathrm{AIII}$ invariant, which vanishes because the origin lies in the topologically trivial far-field point gap. Nevertheless, the sector-resolved invariant may remain nontrivial, for example with \(\nu_+=-\nu_-\neq0\). This is the refined point-gap topology carried specifically by \(z=0\), and it accounts for the symmetry-protected zero-energy boundary state.

Surely, though, such a zero-energy state can be removed by a generic $\mathrm{AIII}$ boundary perturbation of the Hermitian double. If, however, the perturbation preserves the underlying non-Hermitian sublattice symmetry, the two sectors remain decoupled at $z=0$, 
and the corresponding boundary states cannot be removed without closing the bulk gap or breaking the symmetry.

Finally, although protected as a spectral state, the eigenvalue at \(z=0\) carries no finite weight in the normalized thermodynamic-limit spectral density. Everywhere in a sufficiently small neighborhood of the origin, the ordinary point gap is trivial, and the limiting OBC Green’s function therefore coincides with the analytic Brown Green’s function. A nonzero point charge at the origin would contribute a singular term proportional to \(1/z\). As there is no such a term in the Brown (PBC) Green's function (as it does not
support finite DOS weight at the origin), the limiting OBC DOS does not
support finite weight at the origin as well.

The $z=0$ state therefore remains visible in the OBC/SIBC spectrum as a set, while it is absent from the limiting spectral density.

\bibliography{bib}

\end{document}